\documentclass[aps,prc,twocolumn,showpacs,preprintnumbers,nofootinbib,float,superscriptaddress]{revtex4-1}
\usepackage{epsfig}
 
\usepackage[dvipsnames]{xcolor}
\usepackage{float,amsmath,amssymb}
\usepackage{graphicx}
\usepackage{url}
\usepackage{bbold}
\usepackage[utf8]{inputenc}
\usepackage{physics}
\usepackage[caption=false]{subfig}

\newcommand{\xpom}{x_{I\hspace{-0.3em}P}}

\newcommand{\gev}{\mathrm{GeV}}
\newcommand{\Pev}{\mathrm{PeV}}

\newcommand{\Tev}{\mathrm{TeV}}
\newcommand{\bQ}{\overline{Q}}

\newcommand{\xbj}{{x}}
\newcommand{\ygap}{Y_{\rm gap}}

\newcommand{\as}{{\alpha_{\mathrm{s}}}}
\newcommand{\rt}{{\mathbf{r}}}
\newcommand{\ut}{{\mathbf{u}}}

\newcommand{\bt}{{\mathbf{b}}}
\newcommand{\lt}{{\mathbf{l}}}
\newcommand{\Pt}{{\mathbf{P}}}
\newcommand{\Kt}{{\mathbf{K}}}

\newcommand{\kt}{{\mathbf{k}}}

\newcommand{\nc}{{N_\mathrm{c}}}

\newcommand{\fm}{{\rm fm}}

\newcommand{\vk}{\vec{k}}
\newcommand{\vq}{\vec{q}}

\usepackage[breaklinks,colorlinks,citecolor=citcolor,urlcolor=blue,linkcolor=lcolor]{hyperref}
\definecolor{lcolor}{rgb}{0.5,0,0}
\definecolor{citcolor}{rgb}{0,0.3,0.0}
\usepackage[capitalise]{cleveref}
\usepackage{todonotes}

\begin{document}

\title{Inclusive and diffractive neutrino-nucleus scattering at high energy}

\author{Anh Dung Le}

\affiliation{Department of Physics, University of Jyväskylä, P.O. Box 35, 40014 University of Jyväskylä, Finland}
\affiliation{Helsinki Institute of Physics, P.O. Box 64, 00014 University of Helsinki, Finland}

\author{Heikki Mäntysaari}
\affiliation{Department of Physics, University of Jyväskylä, P.O. Box 35, 40014 University of Jyväskylä, Finland}
\affiliation{Helsinki Institute of Physics, P.O. Box 64, 00014 University of Helsinki, Finland}

\begin{abstract}
We calculate the energy dependence of inclusive and diffractive  neutrino-nucleus deep-inelastic scattering cross sections within the dipole picture, focusing on the ultra-high-energy regime. We predict an up to $\sim 10\%$ nuclear suppression in the inclusive neutrino-Oxygen scattering originating from the non-linear QCD dynamics in the small-$x$ Balitsky-Kovchegov evolution. Diffraction is found to be a small $1\dots 4\%$ contribution to the total cross section across a wide range of neutrino energies relevant for current and near-future experiments. The diffractive cross section is calculated separately for the coherent and incoherent channels that are found to be of equal importance. Additionally, we include the dominant contribution from the $|q\bar q g\rangle$ Fock state of the $W^\pm$ and $Z$ bosons in the high-$Q^2$ limit, along with the lowest-order $|q\bar q\rangle$ contribution. The $|q\bar{q}g\rangle$ contribution is found to be numerically significant, reaching up to 40\% of the diffractive cross section.
\end{abstract}

\maketitle

\section{Introduction}

Despite being one of the most abundant particles in the universe, the 
neutrino remains one of the most elusive particles due to its weakly interacting nature. Understanding its properties and interactions with matter is one of the highlights in particle physics, as it can provide hints of BSM physics or reveal insights into the history of the universe, especially in its early stages. Over the years, significant advances have been made in the study of neutrinos in both theoretical and experimental aspects~\cite{SajjadAthar:2021prg,deGouvea:2022gut}. For the latter, various detectors and telescopes have been constructed or designed in order to unveil the puzzles behind these ghost particles.

Neutrinos and their anti-companions cannot be detected directly, but only through the secondary outcomes from their weak interactions with nucleons or nuclei. The measurements of (anti)neutrino events hence rely upon two important energy-dependent factors: the neutrino flux and the neutrino-nucleon or neutrino-nucleus cross section. Therefore, understanding the energy dependence of  (anti)neutrinos scattering off a nucleon or a nucleus is highly demanded. For measurements of neutrino properties, such as the neutrino flavor oscillation, a good knowledge on such scattering observables can significantly reduce the uncertainty~\cite{SajjadAthar:2022pjt}.  

Interaction of (anti)neutrinos with hadronic targets (nucleon, nucleus) is a multi-scale problem, as these projectiles can be emitted from different sources at different energies, and hence are sensitive to different physics. (Anti)-neutrinos can span over a wide energy range, from eV to EeV scales. For neutrino energies $E_{\nu}$ (in the Earth rest frame) below a few MeV, neutrinos interact with nucleus predominantly via elastic scattering or neutrino capture~\cite{Formaggio:2012cpf}. When (anti)neutrinos become more energetic, they can probe more deeply into the internal structure of the nucleon or nucleus. Deep-inelastic scattering (DIS) will then become the dominant process for (anti)neutrinos in the high-energy (HE, $1, \Tev \le E_\nu \le 100\,\Pev$) and ultra-high-energy (UHE, $E_\nu \ge 100 \, \Pev$) regimes, which are the energy ranges considered in this work\footnote{Here we follow the division of energy scales from Ref.~\cite{Ackermann:2022rqc}. 
}. 
The cross section in the HE domain has been measured by the IceCube collaboration~\cite{IceCube:2017roe,Bustamante:2017xuy,IceCube:2020rnc}.
Although the UHE regime has  not been currently reached, there are potential future measurements on this domain at, for example, IceCube-Gen2 detector~\cite{IceCube-Gen2:2020qha}. 

Inspired by progresses and demands of the experimental neutrino physics, numerous theoretical studies \cite{Gandhi:1995tf,Gandhi:1998ri,Kutak:2003bd,Jalilian-Marian:2003ghc,Machado:2003bs,Henley:2005ms,Anchordoqui:2006ta,Ducati:2006vh,Armesto:2007tg,Gluck:2010rw,Connolly:2011vc,Cooper-Sarkar:2011jtt,Chen:2013dza,Goncalves:2014woa,Albacete:2015zra,Arguelles:2015wba,NuSTEC:2017hzk,Reno:2023sdm} have been conducted on the DIS off nucleon (proton) and nucleus of HE and/or UHE (anti)neutrinos, based on different formalisms. A typical approach is to use the collinear factorization and describe the partonic content of the target proton or nucelus in terms of the (nuclear) parton distribution functions. The other approach, that we also employ in this work, is to apply the Color Glass Condensate (CGC)~\cite{Iancu:2003xm} framework to describe the interaction with the dense target as e.g. in Refs.~\cite{Goncalves:2014woa,Ducati:2006vh}.
Most of the current literature on neutrino-nucleon and neutrino-nucleus scatterings focuses on the total inclusive cross-sections. Studies of diffractive scattering are, on the other hand, very limited. They mainly focus on (anti)neutrino-induced diffractive meson and photon production ~\cite{Gaillard:1975nf,Bell:1978ep,Bartl:1977uj,DeLellis:2004ovi}, as they are  important sources of background in the measurement of neutrino oscillation~\cite{NuSTEC:2017hzk}.   
However, to our knowledge, no analyses have been conducted for diffractive dissociation in the hadronic scattering of (anti)neutrinos.  

Although measuring diffractive cross sections in neutrino-target scattering is challenging, it is nevertheless still of interest to investigate diffractive scattering to quantify it for the DIS of above-$\Tev$ (anti)neutrinos, especially for the energies in the UHE domain and with nuclear targets. In such cases, diffractive scattering should be enhanced as the target is close to the black disc limit~\cite{Kowalski:2008sa}.

The current study employs the QCD color dipole model~\cite{Kopeliovich:1981pz,Mueller:1989st,Nikolaev:1990ja} to study the scattering off hadronic targets of (anti)neutrinos in the HE and UHE regimes. This aims at making predictions for the energy-dependence of cross sections for both inclusive and diffractive processes, and off both nucleon and nucleus (in particular, oxygen). The basic ingredient of the calculations is the forward dipole-target elastic amplitude, which at high energies is obtained by solving the nonlinear Balitsky-Kovchegov (BK) equation~\cite{Balitsky:1995ub,Kovchegov:1999yj,Balitsky:2006wa} at small Bjorken-$\xbj$. The non-linear nuclear effects are predicted by constructing the dipole-nucleus scattering amplitude from the dipole-nucleon amplitude using the optical Glauber model at two different stages, either before or after the small-$\xbj$ evolution. This allows us to quantify the significance of the nonlinearities in the nuclear scattering for the HE and UHE (anti)neutrinos.  

The paper is organized as follows. In the next section, we present the dipole model for the (diffractive) deep-inelastic scattering in (anti)neutrino-nucleus scattering and the BK evolution equation. The numerical results for the inclusive scattering are then shown in \cref{sec:inclusive_dis}. In Sec.~\ref{sec:diffraction} we show predictions for diffractive dissociation. We finally present concluding remarks in \cref{sec:conclusions}. Explicit results for the contributions of the $\ket{q\bar q}$ and $\ket{q\bar q g}$ Fock states of the electroweak bosons on the diffractive cross sections are reported in Appendices.~\ref{app:coheren_diffraction} and~\ref{app:incoherent_diffraction}.

\section{Neutrino-nucleus scattering in the dipole picture}

\subsection{Inclusive deep inelastic scattering at high energy}
\label{sec:inclusive_dis}

The scattering of (anti-)neutrino off nucleus can proceed via an exchange of weak vector bosons $W^{\pm}$ or $Z^{0}$ corresponding to charged current (CC) and neutral current (NC) interactions, respectively. The total cross section for neutrino-target scattering can be expressed in terms of the dimensionless structure functions $F_{L;A(N)}(x,Q^2)$, $F_{2;A(N)}(x,Q^2)$ and $F_{3;A(N)}(x,Q^2)$
as follows:
\begin{equation}
\label{eq:incl_diff_xsec}
\begin{aligned}
    \frac{\dd^2\sigma^{CC/NC}_{\nu A(N);\rm tot}}{\dd x\dd Q^2} =  \frac{G_F^2}{4\pi x}&\left(\frac{M_V^2}{M_V^2 + Q^2}\right)^2  \left[ \mathcal{Y}_{+}F_{2;A(N)}^{W^{\pm}/Z} \right. \\
    & \left.- y^2F_{L;A(N)}^{W^{\pm}/Z} \pm \mathcal{Y}_{-}xF_{3;A(N)}^{W^{\pm}/Z}  \right].
\end{aligned}
\end{equation}
The kinematics of the scattering process is completely determined by the Bjorken-$\xbj$ variable and the virtuality of the exchanged vector boson $Q^2$. Here
$y=Q^2/(xs)$ is the inelasticity, $\mathcal{Y}_{\pm} \equiv 1 \pm (1-y)^2$, $G_F$ is the Fermi constant and $M_V$ denotes the mass of the  exchanged vector boson. The squared center-of-mass energy $s$ is related to the neutrino energy $E_\nu$ by $s = 2M_N E_\nu$, where $M_N$ is the nucleon mass. The sign before the left-right asymmetry term $F_3$ depends on the nature of the projectile: it is plus (``$+$") for neutrino, and minus (``$-$") for anti-neutrino. We consider the chiral limit (massless quarks), as large $Q^2$ values are more relevant for neutrino scattering. In this limit, the left-right asymmetry, i.e. the last term in the bracket in \cref{eq:incl_diff_xsec}, is identically zero.

\begin{figure*}[ht!]
    \centering
    \subfloat[Inclusive production \label{subfig:incl}]{%
      \includegraphics[width=0.32\textwidth]{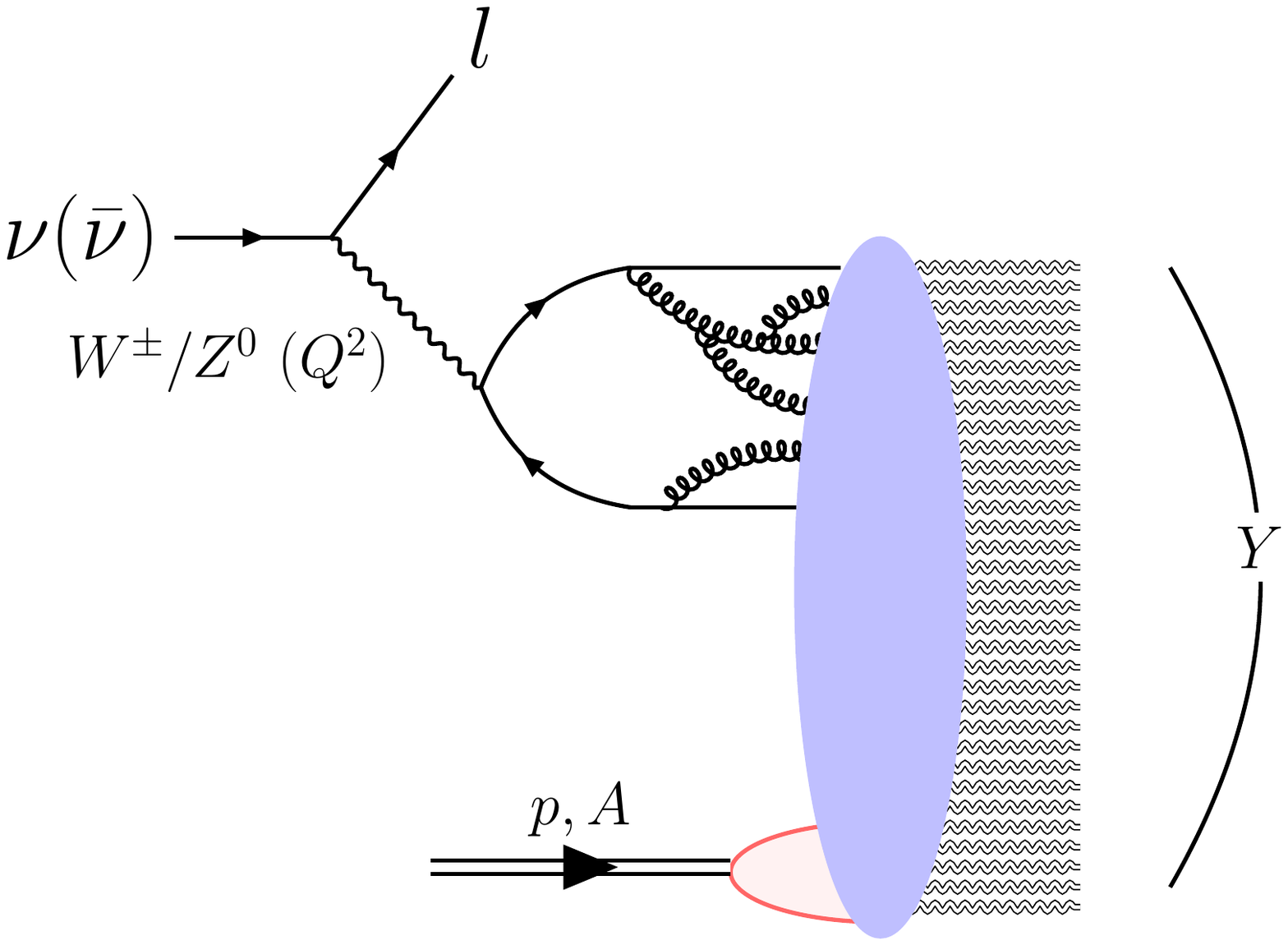}
    }
    \hfill
    \subfloat[Coherent production \label{subfig:coh}]{%
      \includegraphics[width=0.32\textwidth]{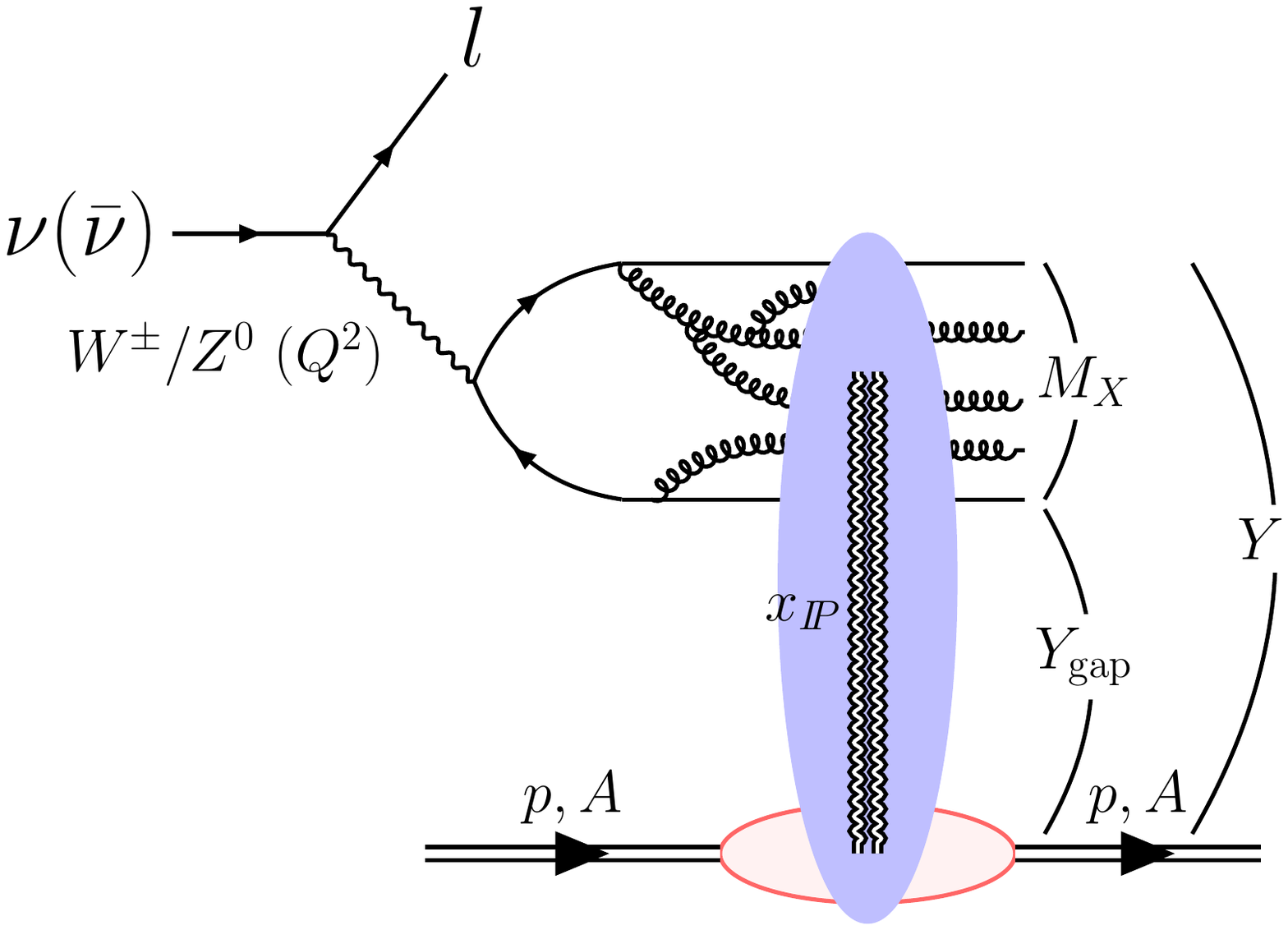}
    }
    \hfill
    \subfloat[Incoherent production \label{subfig:incoh}]{%
      \includegraphics[width=0.32\textwidth]{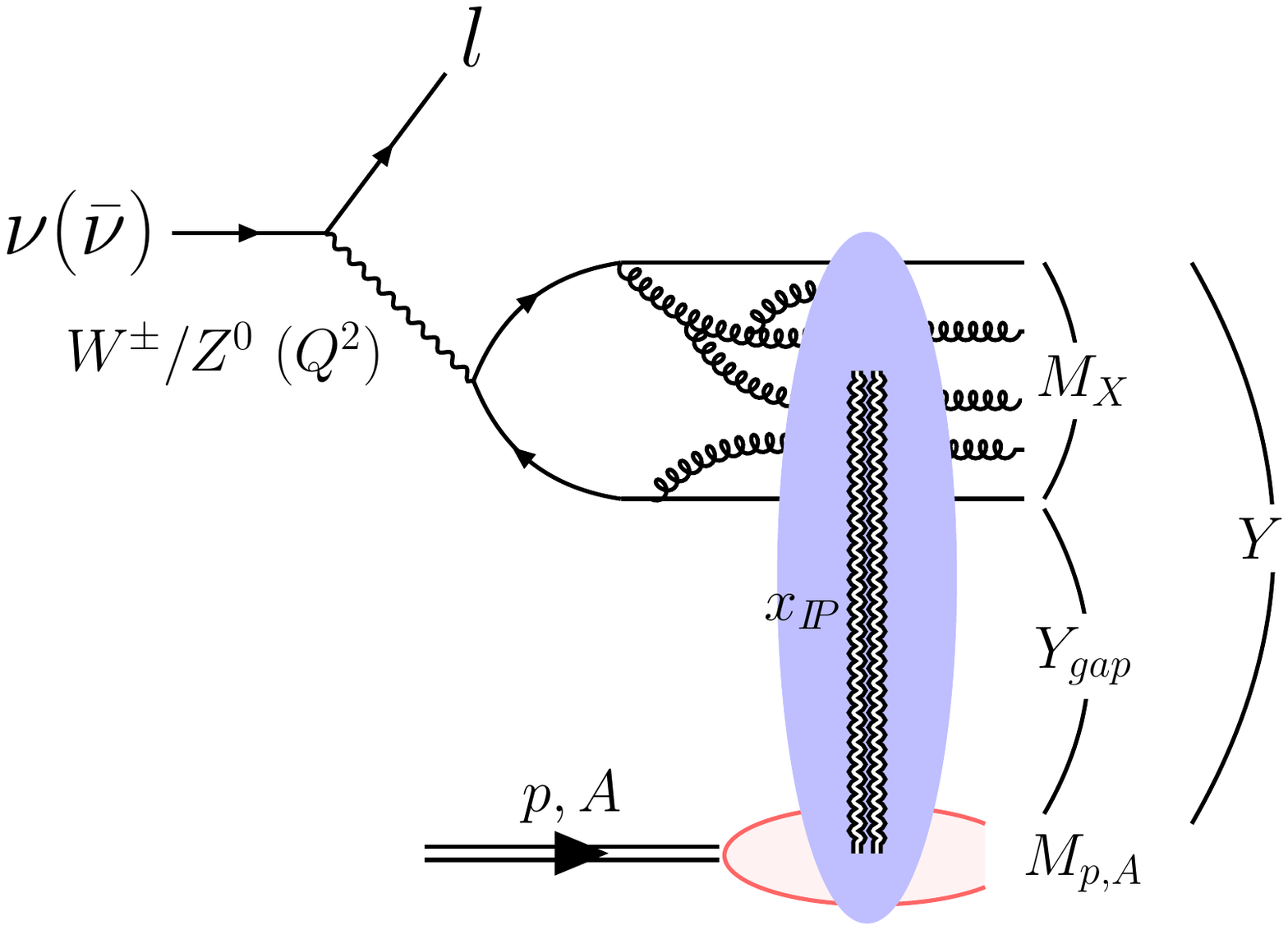}
    }
    \caption{Diagrammatic illustrations of deep-inelastic scattering of (anti-)neutrino off hadronic target in the dipole picture. The inclusive production \protect\subref{subfig:incl} includes any possible final state. For coherent diffraction \protect\subref{subfig:coh}, the target is kept intact, and is separated from the diffractive mass $M_X$ by a rapidity gap. In contrary, the target breaks up in the incoherent diffraction \protect\subref{subfig:incoh}. The rapidity gap in the diffractive production is a consequence of the color-neutral exchange, which is represented by the double wavy lines. 
    }
    \label{fig:DIS}
\end{figure*}

At high energy, the scattering process can be conveniently described in the dipole picture~\cite{Kopeliovich:1981pz,Mueller:1989st,Nikolaev:1990ja} (see \Cref{subfig:incl}). In this framework, at leading order the exchanged vector boson interacts with the target via its quark-antiquark Fock state. Thus the  structure functions can be  expressed in terms of the dipole cross-section $\sigma_\lambda^{CC/NC}$ as
\begin{subequations}
    \begin{equation}
    \begin{aligned}
        F_{L;A(N)}^{W^{\pm}/Z} & = \frac{Q^2}{4\pi^2\alpha_{W/Z}} \sigma^{W^{\pm}/Z}_{\lambda=0},
    \end{aligned} 
    \end{equation}
    \begin{equation}
    \begin{aligned}
        F_{T;A(N)}^{W^{\pm}/Z} & = \frac{Q^2}{4\pi^2\alpha_{W/Z}} \frac{1}{2}\left(\sigma^{W^{\pm}/Z}_{\lambda=+1} + \sigma^{W^{\pm}/Z}_{\lambda=-1}\right).
    \end{aligned} 
    \end{equation}
\end{subequations}
Here $\alpha_{W/Z}$ are coupling constants given in \cref{app:wavefunctions}, $T$ and $L$ refer to the gauge boson transverse or longitudinal polarization states, and the structure function $F_{2;A(N)}^{CC/NC}$ reads
    \begin{equation}
    \label{eq:F2_sf}
    F_{2;A(N)}^{CC/NC} = F_{L;A(N)}^{W^{\pm}/Z} + F_{T;A(N)}^{W^{\pm}/Z}.
    \end{equation}
The dipole cross section can be written as
\begin{equation}
\label{eq:dipole_cross-section}
\begin{aligned}
    \sigma_{\lambda}^{W^{\pm}/Z} = \int \dd^2\rt \int_0^{1} \frac{\dd z}{4\pi z(1-z)} &\left|\Psi^{W^{\pm}/Z}_{\lambda} (\rt, z, Q^2)\right|^2 \\
    &\quad \times\sigma^{q\Bar{q}A(N)}_{\rm tot} (x, \rt).
\end{aligned}
\end{equation}
\Cref{eq:dipole_cross-section} represents the dipole factorization at leading order. Here $\left|\Psi^{V}_{\lambda}(\rt,z,Q^2)\right|^2$ is the squared amplitude for the vector boson $V$ of polarization $\lambda$ and virtuality $Q^2$ to split into a quark-antiquark color dipole of transverse size $\rt$ whose quark leg carries a fraction $z$ of the parent boson's longitudinal momentum. These squared wave functions for $W^\pm$ and $Z^0$ are quoted in \cref{app:wavefunctions}. The cross section for the scattering of a dipole with transverse size $\rt$ is denoted by $\sigma^{q\Bar{q}A(N)}_{\rm tot}(\xbj;\rt)$. 

The dipole-target cross section and the forward elastic dipole-target scattering amplitude $N(\xbj,\rt;\bt)$ are related by the optical theorem 
\begin{equation}
\label{eq:optical_theorem}
    \sigma^{q\Bar{q}A(N)}_{\rm tot} (x, \rt) = 2\int \dd[2]{\bt} N(\xbj,\rt;\bt),
\end{equation}
where $\bt$ is the impact parameter. The small-$\xbj$ evolution of the amplitude $N$ is governed by the Balitsky-Kovchegov (BK) equation~\cite{Kovchegov:1999yj,Balitsky:1995ub}. In the current analysis, we consider the leading-order version involving running-coupling corrections with the Balitsky's prescription~\cite{Balitsky:2006wa}. The corresponding BK equation reads
\begin{widetext}
    \begin{equation}
    \label{eq:bk_equation}
    \partial_Y N(\xbj,\rt;\bt) = \int \dd^2\rt_1 \mathcal{K}_\mathrm{Bal} (\rt,\rt_1,\rt_2) \left[ N(\xbj,\rt_1;\bt_1) + N(\xbj,\rt_2;\bt_2) - N(\xbj,\rt;\bt) - N(\xbj,\rt_1;\bt_1)N(\xbj,\rt_2;\bt_2)\right] 
    \end{equation}
\end{widetext}
where $\rt_2 = \rt - \rt_1$, $\bt_1 = \bt - \rt_2/2$, $\bt_2 = \bt + \rt_1/2$. The kinematical variable $Y$ denotes the evolution rapidity defined as $Y\equiv\ln(1/x)$. The  kernel $\mathcal{K}_{Bal}$ in \cref{eq:bk_equation} reads
\begin{equation}
    \label{eq:Balitsky_kernel}
    \begin{aligned}
        \mathcal{K}_\mathrm{Bal}(\rt,\rt_1,\rt_2) = \frac{\alpha_s(\rt^2)}{2\pi} &\left[\frac{\rt^2}{\rt_1^2\rt_2^2} + \frac{1}{\rt_1^2}\left(\frac{\alpha_s(\rt_1^2)}{\alpha_s(\rt_2^2)}-1\right)\right.\\
    &\left. + \frac{1}{\rt_2^2}\left(\frac{\alpha_s(\rt_2^2)}{\alpha_s(\rt_1^2)}-1\right) \right].
    \end{aligned}
\end{equation}
The running coupling in the above formula is given by
\begin{equation}
\label{eq:running-coupling}
    \alpha_s(\rt^2) = \frac{12\pi}{(11\nc-2N_f)\ln\frac{4C^2}{\rt^2\Lambda_\mathrm{QCD}^2}},
\end{equation}
where $\nc=3$, and $\Lambda_\mathrm{QCD}$ is fixed to the value $0.241~\gev$. The running coupling scale $C^2$ is determined by the applied DIS fit \cite{Lappi:2013zma}. As only light-quark contribution is included in the applied fit and in this work, we set $N_f=3$. Furthermore, to avoid the Landau pole, $\as$ is frozen at the value $0.7$
in the infrared. 

Although the BK equation with an impact parameter dependence can be solved numerically~\cite{Berger:2010sh}, it is computationally demanding, and the total cross section derived from its solution does not respect the unitary Froissart-Martin bound due to the presence of long-range Coulomb fields away from the saturation region~\cite{Kovner:2001bh}. As such it would require one to additionally introduce a model for confinement scale physics as e.g. in Refs.~\cite{Berger:2012wx,Kovner:2001bh,Schlichting:2014ipa} (see also related developments in Ref.~\cite{Cepila:2018faq}). 
Instead of the full $\bt$-dependence, we consider simpler assumptions. For a nucleon (proton) target, the amplitude is assumed to factorize as:
\begin{equation}
\label{eq:proton_factorization}
    N(\xbj,\rt;\bt) = T_p(\bt)\mathcal{N}(\xbj,\rt),
\end{equation}
where the impact-parameter profile $T_p$ is normalized as
\begin{equation}
\label{eq:norm_Tp}
    \int \dd^2\bt~ T_p(\bt) = \frac{\sigma_0}{2}.
\end{equation}
The BK equation (\ref{eq:bk_equation}) is then reduced to become an equation for the $\bt$-independent amplitude $\mathcal{N}$. It requires as an input the dipole-target scattering amplitude at some initial $\xbj=\xbj_0$, which can be chosen as
\begin{equation}
\label{eq:init_bk_proton}
    \mathcal{N}(\xbj=x_0,\rt) = 1 - \exp\left[-\frac{\rt^2Q_{s,0}^2}{2}\ln \left(e\cdot e_c+\frac{1}{|\rt|\Lambda_{QCD}}\right)\right],
\end{equation}
based on the McLerran-Venugopalan (MV)~\cite{McLerran:1993ni} model. Here the initial condition is parametrized at $x_0=0.01$. The free parameters $Q_{s,0}^2$ and $e_c$ in \cref{eq:init_bk_proton} can be obtained from fits to inclusive structure function data~\cite{Casuga:2023dcf,Lappi:2013zma,Albacete:2010sy,Hanninen:2022gje,Beuf:2020dxl}. In this work we use two fits from Ref.~\cite{Lappi:2013zma}, which are referred to as MV ($e_c=1$ fixed) and MV$^e$ ($e_c$ free), in order to estimate the uncertainty originating from these fits (see also Ref.~\cite{Casuga:2023dcf} for a recent determination of the BK initial condition with uncertainty estimates showing that the DIS data constrain the initial condition accurately). These two fits also determine the corresponding values of the normalization factor $\sigma_0/2$ and the running coupling scale $C^2$ introduced above. 

Generalization to nuclear targets can be done via two different setups. In the first case, the initial dipole-nucleon amplitude \eqref{eq:init_bk_proton} is generalized to the dipole-nucleus case at the initial condition of the BK evolution following Refs.~\cite{Lappi:2013zma,Kowalski:2003hm} but without the large-$A$ limit:
\begin{equation}
\label{eq:before_evol}
            N_A(x=\xbj_0,\rt;\bt) = 1 - \left[1 - \frac{\sigma_0}{2}T_A(\bt)\mathcal{N}(\xbj_0,\rt)\right]^A.
\end{equation}
Here $T_A(\bt)$ is the Woods-Saxon density profile given by
\begin{equation}
\label{eq:Woods-Saxon}
    T_A(\bt) = \int_{-\infty}^{\infty} \dd z \frac{\rho_0}{1+ \exp\left[\frac{\sqrt{z^2+\bt^2}-R_A^2}{d}\right]}
\end{equation}
with $d= 0.54~ \fm, R_A = (1.12A^{1/3} - 0.86 A^{-1/3})~\fm$, and $\rho_0$ being the normalization factor such that $\int \dd^2\bt T_A(\bt) = 1$. Note that we choose a different normalizations (and units) for the proton and nuclear density profiles $T_A$ and $T_p$. Although more realistic nucleon density distributions are available~\cite{Carlson:1997qn}, we expect the simple Woods-Saxon distribution to be suitable in this work when calculating the nuclear effects that are not sensitive to the detailed nuclear geometry.
The dipole amplitudes at different impact parameters $|\bt|$ are then evolved independently of each other. We will refer to this setup as \emph{before evol.} later in this work. 

For comparison, we also consider the case where we first evolve the dipole-nucleon (proton) amplitude from the initial $x_0$ to the desired Bjorken-$\xbj$, and then construct the dipole-nucleus amplitude by using the evolved dipole-proton amplitudes $\mathcal{N}(\xbj,\rt)$ in Eq.~\eqref{eq:before_evol}. We will refer to this as \emph{after evol}. This allows us to quantify the importance of saturation effects in the small-$x$ evolution due to the reason that the saturation scale of the evolving nucleus is higher than that of the nucleon. Furthermore, this setup is necessary when we calculate diffractive cross section with an event-by-event fluctuating target geometry, as we want to avoid solving the impact parameter dependent BK evolution in this work.

Two approaches which are quite similar to our settings were employed in Refs.~\cite{Cepila:2020xol,Bendova:2020hbb} to construct the nuclear amplitudes. The first approach, referred to as b-BK-GG, is to couple the dipole-proton amplitude obtained as a solution to the BK equation to the Glauber-Gribov formalism. This is similar to our \emph{after evol.} setting. Alternatively, one can solve directly the BK equation for the dipole-nucleus amplitude. This so-called b-BK-A approach is rather similar to the above \emph{before evol.} prescription. In the context of photoproduction of vector meson ($\mathrm{J}/\psi$) in ultra-peripheral Pb-Pb collisions~\cite{Bendova:2020hbb}, the LHC Run 1 data ($\sqrt{s} = 2.76~\Tev$) prefers the b-BK-A approach. It was also shown in Ref.~\cite{Cepila:2020xol} to provide better descriptions to the Fermilab E665 data on nuclear suppression for both Ca and Pb nuclei. 

In this work we consider Oxygen nuclei, as that can be taken to be a good estimate for the earth atmosphere. In addition, Oxygen is also phenomenologically relevant for some experiments such as IceCube, for which neutrinos can collide with nuclei inside water molecules. 

We shall focus on the energy-dependence of the integrated cross-sections defined by
\begin{equation}
\label{eq:integratED_xsec}
    \sigma_{\nu A(N);\rm tot}^{CC/NC}(E_{\nu}) = \int^{\xbj_{\rm max}s}_{Q_{\rm min}^2}\dd Q^2 \int_{Q^2/s}^{\xbj_{\rm max}} \dd x \frac{\dd^2\sigma^{CC/NC}_{\nu A;\rm tot}}{\dd x\dd Q^2},
\end{equation}
where the virtuality cut-off $Q_{\rm min}^2$ is fixed at $1~\gev^2$. The value of the longitudinal cut-off $\xbj_{\rm max}$ defines the scenario in consideration. If $\xbj_{\rm max}=\xbj_0$, only the small-$\xbj$ sector is involved in the calculation. Otherwise, if $\xbj_{\rm max}=1$, the large-$\xbj$ contribution is also taken into account. In the latter case, the dipole-nucleon amplitude is extrapolated to the large-$\xbj$ regime ($\xbj > \xbj_0$) as
\begin{equation}
\label{eq:large-x-amp}
    \mathcal{N}(\xbj,\rt) = \mathcal{N}(\xbj_0,\rt) \left(\frac{1-\xbj}{1-\xbj_0} \right)^{6}.
\end{equation}
This effectively suppresses the dipole amplitude in the large-$x>x_0$ region (where the applied framework is not applicable) similarly as in Ref.~\cite{Goncalves:2013kva} in the context of neutrino-DIS in dipole picure, as well as in the phenomenologically succesfull IPsat parametrization for the dipole-proton scattering~\cite{Kowalski:2003hm}. We will quantify the sensitivity of our results on this extrapolation in the high-$\xbj$ domain in Sec.~\ref{sec:result_inclusive}.

\subsection{Diffractive deep inelastic scattering}

One striking phenomenon in the deep inelastic scattering of lepton off a hadronic target is hard diffraction. This is characterized by the existence of a rapidity gap $\ygap$ (no particles present) between the diffractively produced   system and the final state of the target (see \Cref{subfig:coh,subfig:incoh}). This rapidity gap is the signature of the color-singlet exchange. This exchange involves two gluons at lowest order in perturbative QCD, which makes diffraction more sensitive to the gluonic content inside the target than the inclusive production. In analogy to the inclusive cross section shown in \cref{eq:incl_diff_xsec}, the diffractive cross sections in the chiral limit are given in terms of diffractive structure functions as 
\begin{equation}
\label{eq:dfrac_diff_xsec}
\begin{aligned}
    &\frac{\dd^3\sigma_{\nu A (N);D}^{CC/NC}}{\dd\xpom \dd\xbj \dd Q^2} = \frac{G_F^2}{4\pi\xbj} \left(\frac{M_V^2}{M_V^2 + Q^2}\right)^2 \\
    &\qquad\quad \times\left(\mathcal{Y}_{+}F_{2;A(N)}^{D(3);W^{\pm}/Z} - y^2 F_{L;A(N)}^{D(3);W^{\pm}/Z} \right).
\end{aligned}
\end{equation}
Here $\xpom$ controls the size of the rapidity gap, $\ygap = \ln(1/\xpom)$, within the total rapidity interval, $Y = \ln(1/\xbj)$. 
Note that in the case of a diffractive scattering, three invariants are required to describe the kinematics, in this case we choose to use $\xpom,x$ and $Q^2$. The diffractive structure functions depend on all these variables. The mass of the diffractively produced system can be written as $M_X^2 = Q^2((\xpom/x) - 1)$.

Diffractive production can be coherent (\Cref{subfig:coh}) or incoherent (\Cref{subfig:incoh}) depending on the nature of the target's final state. For coherent interaction, the target remains in its ground state after the scattering. This requires the average over configurations of the target's wave function at the amplitude level, which schematically reads~\cite{Good:1960ba}
\begin{equation}
\label{eq:coherent_diffraction}
    \sigma_{\nu A (N);D}^{CC/NC}\bigg|_{\rm coh} \propto \left|\left<\mathcal{A}\right>\right|^2,
\end{equation}
where $\mathcal{A}$ is the diffractive scattering amplitude and $\langle \mathcal{O} \rangle$ denotes the average over possible target configurations. For a detailed definition of the scattering amplitude, see e.g. Refs.~\cite{Beuf:2022kyp,Beuf:2024msh}.
In other words, coherent diffraction probes the average dipole-target interaction, i.e. the average spatial distribution of gluons in the target.

The target can also break up, which corresponds to the incoherent production. This can be calculated by averaging over configurations at the cross-section (squared amplitude) level and subtracting the coherent (no breakup) part. The incoherent diffractive cross sections are then given by the variance of the amplitude~\cite{Miettinen:1978jb,Caldwell:2010zza} 
\begin{equation}
\label{eq:incoherent_diffraction}
    \sigma_{\nu A (N);D}^{CC/NC}\bigg|_{\rm incoh} \propto  \big<\left|\mathcal{A}\right|^2\big > - \left|\big<\mathcal{A}\big>\right|^2.
\end{equation}
Consequently, it can measure the fluctuations in the wave function of the target~\cite{Mantysaari:2020axf,Mantysaari:2016ykx}. The first term in \cref{eq:incoherent_diffraction} can be obtained by computing the diffractive structure functions event-by-event and then performing the average over configurations. 
The event-by-event quantum fluctuations of the nuclear wave functions can originate from several sources. One important origin is random distribution of nucleons within the nucleus (according to the Woods-Saxon distribution). In addition, there exists  fluctuations in the color charge distribution inside nucleons, and potentially fluctuations in the overall density characterized by the saturation scale $Q_s$. Furthermore, nucleon geometry can also fluctuate event-by-event~\cite{Mantysaari:2016ykx,Mantysaari:2020axf}. We only take into account  fluctuating nucleon positions in this analysis, but note that  nucleon substructure fluctuations have been shown to increase the incoherent cross section in $\gamma+\mathrm{Pb}$ scattering by approximatively 50\%~\cite{Sambasivam:2019gdd,Mantysaari:2022sux} in the case of $\mathrm{J}/\psi$ production. For this purpose, the event-by-event transverse positions for nucleons are sampled from the Woods-Saxon distribution given in \cref{eq:Woods-Saxon}. The averages in \cref{eq:coherent_diffraction,eq:incoherent_diffraction} are then given by
\begin{subequations}
\label{eq:target_average}
    \begin{equation}
    \left< \mathcal{O} \right> = \int \left[ \prod_i^A \dd^2\bt_i T_A(\bt_i) \right] \mathcal{O}, 
    \end{equation}
    and
    \begin{equation}
    \left< |\mathcal{O}|^2 \right> = \int \left[ \prod_i^A \dd^2\bt_i \dd^2\bt'_i T_A(\bt_i) T_A(\bt'_i) \right] \mathcal{O}^{\dagger}\mathcal{O}, 
    \end{equation}
\end{subequations}
where $\bt_i$ ($\bt'_i$) is the transverse position of the $i^{th}$ nucleon in the amplitude (conjugate amplitude). 

The diffractive system results from the hadronization of the Fock state of the vector boson $V$ involved in the strong interaction. The diffractive structure functions in \cref{eq:dfrac_diff_xsec} then include contributions from infinitely many Fock states of the relevant vector boson $V$. Since the scattering is dominated by the large-$Q^2$ regime, one expects that the two lowest-order states $\ket{q\Bar{q}}$ and $\ket{q\Bar{q}g}$ should give the most important contribution to the cross-sections. Schematically, their contributions are proportional to the squared dipole amplitudes in the fundamental and adjoint representations, respectively. The diffractive $q \bar q g$ production in photon-nucleus collisions in exact kinematics is known~\cite{Beuf:2022kyp,Beuf:2024msh}. However, at large $Q^2$, only the transverse component of the $q\bar q g$ contribution is relevant, as it is enhanced by the large transverse logarithm $\log Q^2$~\cite{Wusthoff:1997fz,Golec-Biernat:1999qor,Kowalski:2008sa} (the so called ``W\"usthoff result''). The diffractive $q\Bar{q}$ contribution and the $q\Bar{q}g$ component in the large-$Q^2$ limit calculated in this work are given in \cref{app:coheren_diffraction,app:incoherent_diffraction}. Both components are included in our numerical results. Specifically we use \cref{seq:F_qq} to calculate the $q\bar q$ contribution to the coherent cross section, and \cref{eq:qqg_diff_structurefun} for the $q\bar q g$ contribution. The corresponding results for the total diffractive cross section (sum of coherent and incoherent) are given by \cref{eq:totaldiff_qq_W_Z,eq:totaldiff_qqg_W_Z}.

From \cref{eq:coherent_diffraction}, it is also worth noting that the coherent diffractive cross-section for the neutrino-nucleon (proton) scattering is proportional to the squared impact-parameter profile $T_p^2(\bt)$ when the impact parameter profile is assumed to factorize as  in \cref{eq:proton_factorization}:
\begin{equation}
\label{eq:}
    \sigma_{\nu N;D}^{CC/NC}  \propto \int \dd[2]\bt T_p^2(\bt).
\end{equation}
Consequently, while the inclusive cross-sections are not sensitive to the detailed form of $T_p$ as long as the normalization of $T_p$ in \cref{eq:norm_Tp} is fixed, the diffractive cross-sections depend strongly on the proton (nucleon) shape. Here we use the incomplete gamma function profile to parametrize the proton density profile:
\begin{equation}
    T_p(\bt) = \frac{\Gamma\left(\frac{1}{\omega},\frac{\pi \bt^2}{(\sigma_0/2)\omega}\right)}{\Gamma\left(\frac{1}{\omega}\right)}.
\end{equation}
Here the parameter $\omega$ controls the shape of the profile. Its optimal values ($\omega=1.24$ and   $\omega=2.32$ for the MV and MV$^e$ parametrizations) for the applied DIS fits were obtained from a fit to the HERA diffractive cross section data~\cite{H1:2012xlc} in the low mass region in Ref.~\cite{Lappi:2023frf}. 

As in the inclusive case, we only consider integrated cross sections. In particular, we calculate
\begin{equation}
\label{eq:integratED_dfrac_xsec}
    \sigma_{\nu A (N);D}^{CC/NC}(E_\nu) = \int\limits_{Q_{min}^2}^{x_{\rm max}s} \dd Q^2 \int\limits_{Q^2/s}^{x_{\rm max}} \dd \xbj\int\limits_x^{x_{\rm max}} \dd\xpom  \frac{\dd^3\sigma_{\nu A;A}^{CC/NC}}{ \dd\xpom \dd\xbj \dd Q^2}. 
\end{equation}
Here, the condition $\xpom\ge\xbj$ is imposed as the rapidity gap can never exceed the total available rapidity.  

\section{Inclusive scattering}
\label{sec:result_inclusive}

\begin{figure*}[tb]
    \centering
    \includegraphics[width=\textwidth]{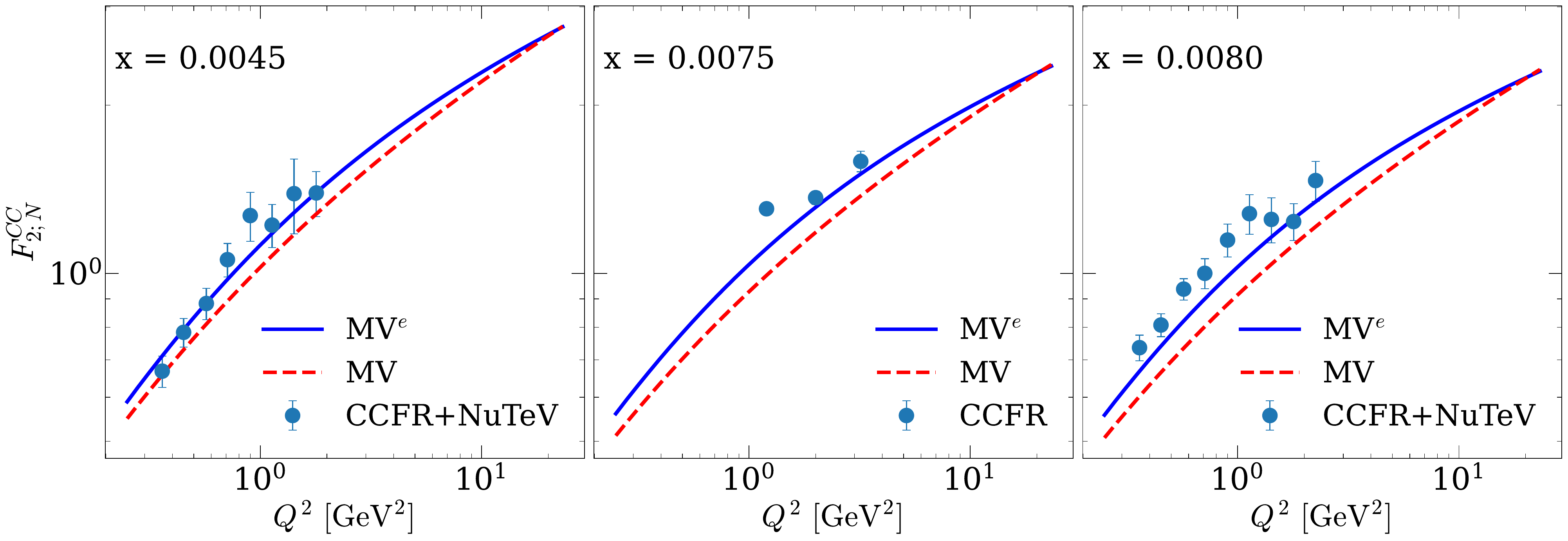}
    \caption{DIS structure function $F_{2;N}^{CC}$ at small-$\xbj$ for the scattering off nucleon target and charged-current exchange. Results are compared with the CCFR~\cite{CCFR:1997tam} ($x=0.0075$) and CCFR+NuTeV~\cite{CCFR:2000ihu} ($x=0.0045, x=0.008$) data.
    }
    \label{fig:F2N_CC}
\end{figure*}
We start by comparing the  dipole model predictions for the structure function $F_{2;N}^{CC}$ at small $x$ with the available CCFR and NuTeV data from  Refs.~\cite{CCFR:1997tam,CCFR:2000ihu}. The comparison is presented in \cref{fig:F2N_CC}. 
These measurements cover kinematics where the neutrino energy is below the HE domain. However, this comparison can serve as a test for our setup, as all free parameters in the applied setup are obtained from fits to independent HERA electron-proton scattering measurements. The calculations using the MV$^{e}$ initial condition of Ref.~\cite{Lappi:2013zma} are seen to describe the data better than the ones using the MV parametrization. In addition, a better description of the data is achieved at the smallest $\xbj$. This is because at higher $\xbj$, the valence-quark contribution, which is not included in our calculation, becomes more important. As shown in Ref.~\cite{Ducati:2006vh}, although this valence component is subdominant at small-$\xbj$, it can still have a numerically non-negligible effect at $\xbj \sim10^{-2}$. 
Overall, the numerical calculation provides an appropriate description of the available structure function data at small-$\xbj$. It hence suggests that the calculation based on our setup can produce realiable predictions for the neutrino DIS in the HE and UHE regimes given that the energy ($\xbj$) dependence is obtained by solving the perturbative BK equation. 

\begin{figure*}[tb]
\centering
    \includegraphics[width=\textwidth]{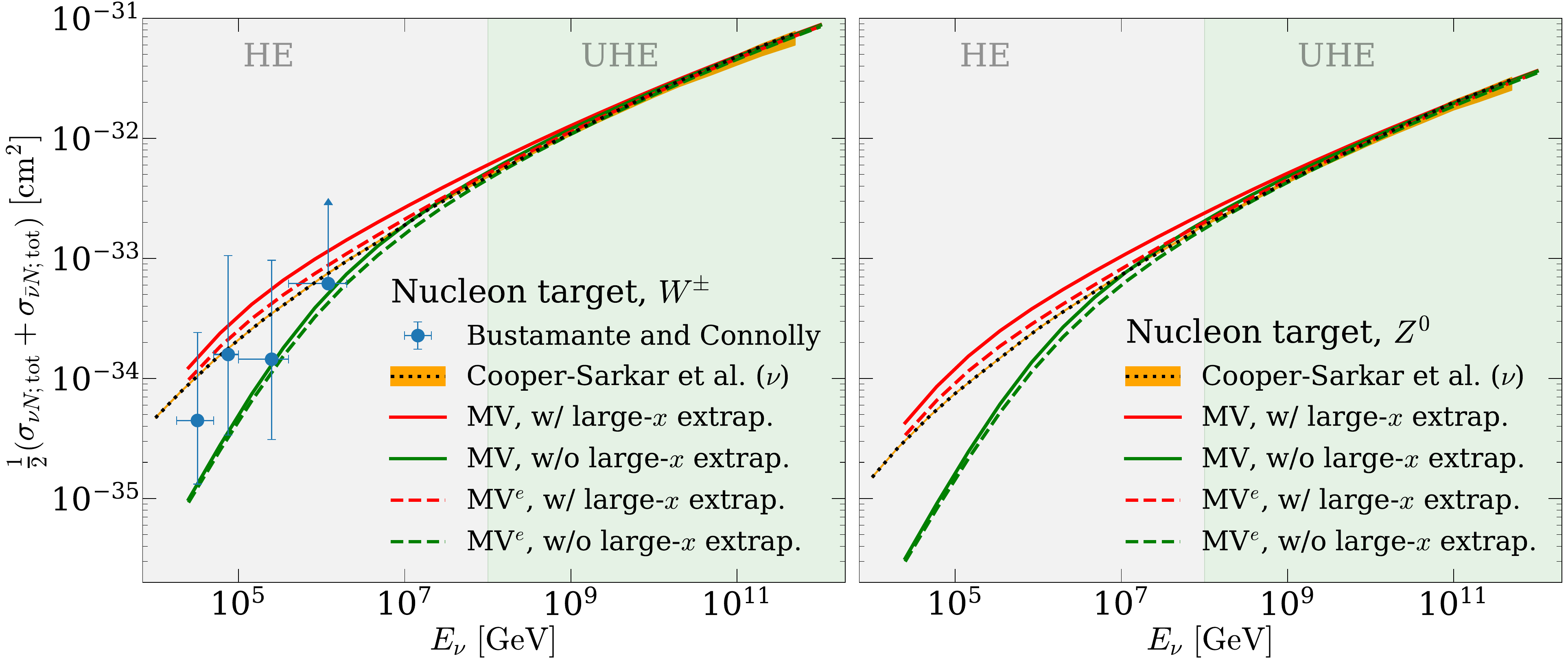}
    \caption{Inclusive neutrino-nucleon cross-sections for charged current ($W^{\pm}$, left panel) and neutral current ($Z^0$, right panel) exchanges. The ``Cooper-Sarkar et al.'' curves represent the DGLAP-based result from Ref.~\cite{Cooper-Sarkar:2011jtt}. The blue points are extracted from the IceCube data in Ref.~\cite{Bustamante:2017xuy}. 
    Results are shown with and without the large-$x$ extrapolation.
    } 
    \label{fig:inclusive_proton}
\end{figure*}

The total (inclusive) neutrino-proton cross section for neutrino energies covering the high-energy (HE) and ultra-high-energy (UHE) regimes are shown in \Cref{fig:inclusive_proton}. The cross sections rise towards higher neutrino energies, and increase almost linearly in the logarithmic scale for the UHE neutrino, suggesting that there is no significant saturation effect visible. Furthermore, larger cross sections are seen for the charged current than for the neutral current. In particular, the former is roughly three times larger than the latter, which is the common expectation from the literature. 

The results are shown separately with and without the large-$x$ extrapolation, i.e. by setting $x_\mathrm{max}=0.01$ and using \cref{eq:large-x-amp} at $x>x_\mathrm{max}$, or setting $x_\mathrm{max}=1$, respectively. The total cross section is found to depend strongly on this choice at energies below $\sim 10^7\,\gev$. This strong dependence limits the predictive power of our calculation in that region. The results obtained using both extrapolation schemes are compatible with the available IceCube data extracted in Ref.~\cite{Bustamante:2017xuy}.
Meanwhile the cross-sections in the UHE regime are shown to be sensitive only to the small-$\xbj$ regime and as such considered to be more reliable. From now on, only the results with the large-$x$ extrapolation are shown.

Especially in the UHE regime the difference between the different fits is small, 
the MV$^e$ fit resulting in slightly lower cross section. With this fit the cross section is also closer to the results from the collinear factorization approach obtained using the NLO DGLAP evolution, reported in Ref.~\cite{Cooper-Sarkar:2011jtt} and also shown in \cref{fig:inclusive_proton} (denoted by \emph{Cooper-Sarkar et al.}). In the UHE regime one could see a slight suppression in the rise of cross sections calculated using the CGC approach of the current work compared to the conventional collinear approach. Fitting the parametrization $\sigma_{\nu N} \propto (E_\nu/1~\gev)^{\alpha}$ to the  \emph{Cooper-Sarkar et al.} data and to our numerical results for $E_\nu\ge 10^{7}~\gev$ (as done in \cite{Formaggio:2012cpf}), one gets the following results for the charge-current exchange:
\begin{align}
    \sigma^{CC}_{\nu N; CS} &\simeq \left( 9.222\times10^{-36} {\rm~ cm^2}\right) \left(\frac{E_\nu}{1~\gev}\right)^{0.339},  \\
    \sigma^{CC}_{\nu N; {\rm MV}^e} &\simeq \left( 1.426\times10^{-35} {\rm~ cm^2}\right) \left(\frac{E_\nu}{1~\gev}\right)^{0.319}, \label{eq:edep_param_proton_mve} \\
    \sigma^{CC}_{\nu N; {\rm MV}} &\simeq \left( 2.281\times10^{-35} {\rm~ cm^2}\right) \left(\frac{E_\nu}{1~\gev}\right)^{0.302}  \label{eq:edep_param_proton_mv} .
\end{align}
Here, the energy-evolution slope from \emph{Cooper-Sarkar et al.} (CS) is slightly larger than those from our numerical results, that can be seen as a rather weak hint of a saturation effect.
Similar but a more strong suppression in the BK evolved cross sections has been obtained previously in Ref.~\cite{Albacete:2015zra}, especially if a subset of higher order corrections is included in the evolution.
This observation indicates that the gluon saturation effect can become relevant at highest possible neutrino energies. 

The MV$^e$ fit  results in a better agreement with the IceCube data~\cite{Bustamante:2017xuy} in the HE regime compared to the MV parametrization. Technically this is because this initial condition is a more steep function of $r$ at 
$r\sim 1/Q_{s,0}$ due to the infrared regulator parameter $e_c$ in \cref{eq:init_bk_proton}. As suggested in Ref.~\cite{Lappi:2013zma}, such steep initial condition can provide a better effective description of the high-$Q^2$  DGLAP dynamics not fully included in the small-$x$ framework employed here. In fact, as shown in Refs.~\cite{Lappi:2013zma, Lappi:2023frf}, both inclusive and diffractive HERA data generally favor the MV$^{e}$ input. 
The good agreement with both the $F^{CC}_{2;N}$ and neutrino-nucleus cross section data demonstrates that especially the MV$^e$ fit can be used to obtain reliable estimates for neutrino-DIS cross sections.

\begin{figure*}[tb]
    \centering
    \includegraphics[width=\textwidth]{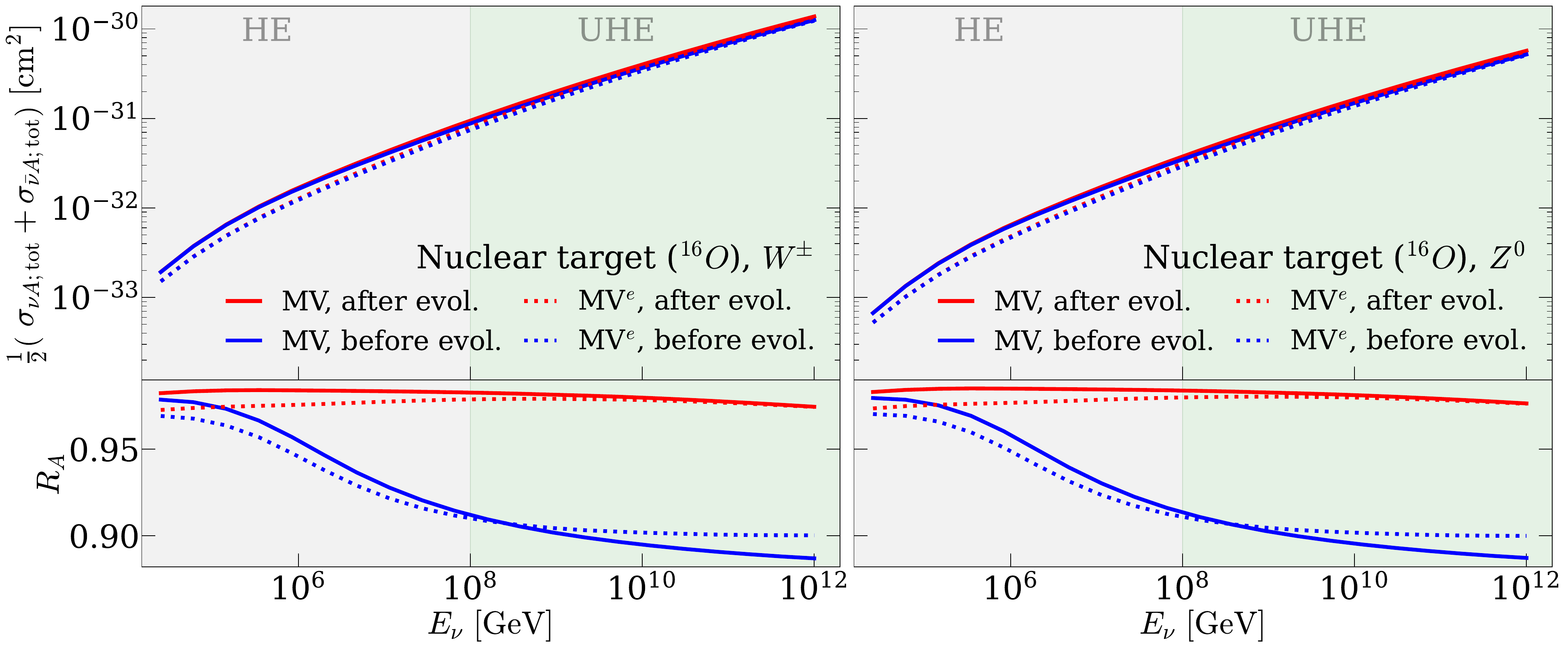}
    \caption{Inclusive neutrino-oxygen cross-sections for charged current ($W^{\pm}$, left panel) and neutral current ($Z^0$, right panel). Numerics for the finite-$A$ setup both after and before the evolution are shown. 
    }
    \label{fig:inclusive_nucl}
\end{figure*}

Predictions for the inclusive cross section with a nuclear target ($^{16}O$) are shown in \Cref{fig:inclusive_nucl}, together with the nuclear modification ratio defined as
\begin{equation}
\label{eq:nucl_mod_ratio}
    R_A = \frac{\sigma_{\nu A;\mathrm{tot}}}{A\sigma_{\nu N;\mathrm{tot}}}.
\end{equation}
We again show results using both the MV and MV$^e$ fits. Furthermore, we compare cross sections calculated by generalizing the dipole-proton amplitude to the dipole-nucleus case before or after the small-$x$ BK evolution as discussed in Sec.~\ref{sec:inclusive_dis}.

The energy dependence of the neutrino-nucleus cross sections is shown in the upper panels of \cref{fig:inclusive_nucl}. We again find a slightly smaller cross section in the HE region when the MV$^e$ fit is used, but overall the dependence on the applied fit, as well as on the method chosen to generalize the dipole-nucleon amplitude to the dipole-nucleus case, is small. The energy dependence of the calculated netrino-nucleus cross section    
at $E_\nu \ge 10^7 ~\gev $ can be approximated by the following power-law expressions:
\begin{align}
    \sigma^{CC(a)}_{\nu A; {\rm MV}^e} &\simeq \left( 2.245\times10^{-34} {\rm~ cm^2}\right) \left(\frac{E_\nu}{1~\gev}\right)^{0.319}, \\
    \sigma^{CC(a)}_{\nu A; {\rm MV}} &\simeq \left( 3.639\times10^{-34} {\rm~ cm^2}\right) \left(\frac{E_\nu}{1~\gev}\right)^{0.302},
\end{align}
for the {\em after evol.} setup, and
\begin{align}
    \sigma^{CC(b)}_{\nu A; {\rm MV}^e} &\simeq \left( 2.147\times10^{-34} {\rm~ cm^2}\right) \left(\frac{E_\nu}{1~\gev}\right)^{0.317}, \\
    \sigma^{CC(b)}_{\nu A; {\rm MV}} &\simeq \left( 3.567\times10^{-34} {\rm~ cm^2}\right) \left(\frac{E_\nu}{1~\gev}\right)^{0.299},
\end{align}
for the {\em before evol.} setup. 

In the {\em befor evol.} setup the energy dependence is slightly slower than in the neutrino-nucleus scattering, see Eqs.~\eqref{eq:edep_param_proton_mve} and \eqref{eq:edep_param_proton_mv}. This is because the non-linear term in the BK equation slows down the evolution more for denser systems, implying that a moderate saturation effect can be seen in the UHE regime. In contrast, in the {\em after evol.} setup no modification to the energy dependence is visible, as the small-$x$ evolution is solved for independent nucleons. 

The nuclear suppression (\cref{fig:inclusive_nucl}, lower panel) is found to be in the $2\dots 5$\% range in the HE region covered by current data in all setups. In the UHE domain, up to $\sim 10\%$ suppression can be reached, and very similar results are obtained using both the MV$^e$ and MV fits. These results correspond to the more realistic setup where we evolve the nucleus constructed at $x=x_0$ (\emph{before evol.} setup). On the other hand, if we evolve individual nucleons (\emph{after evol.} setup), we obtain a nuclear suppression factor $R_A$ that has a much weaker energy dependence and that is much closer to unity. As already seen above when comparing the energy dependence of the cross section, this noncommutativity between the nuclear setup and the high-energy evolution shows that the total neutrino-nucleus cross sections in the UHE region are sensitive to the saturation effects in the small-$x$ evolution. 

\section{Diffractive scattering}
\label{sec:diffraction}

\begin{figure*}[ht!]
    \centering
    \includegraphics[width=\textwidth]{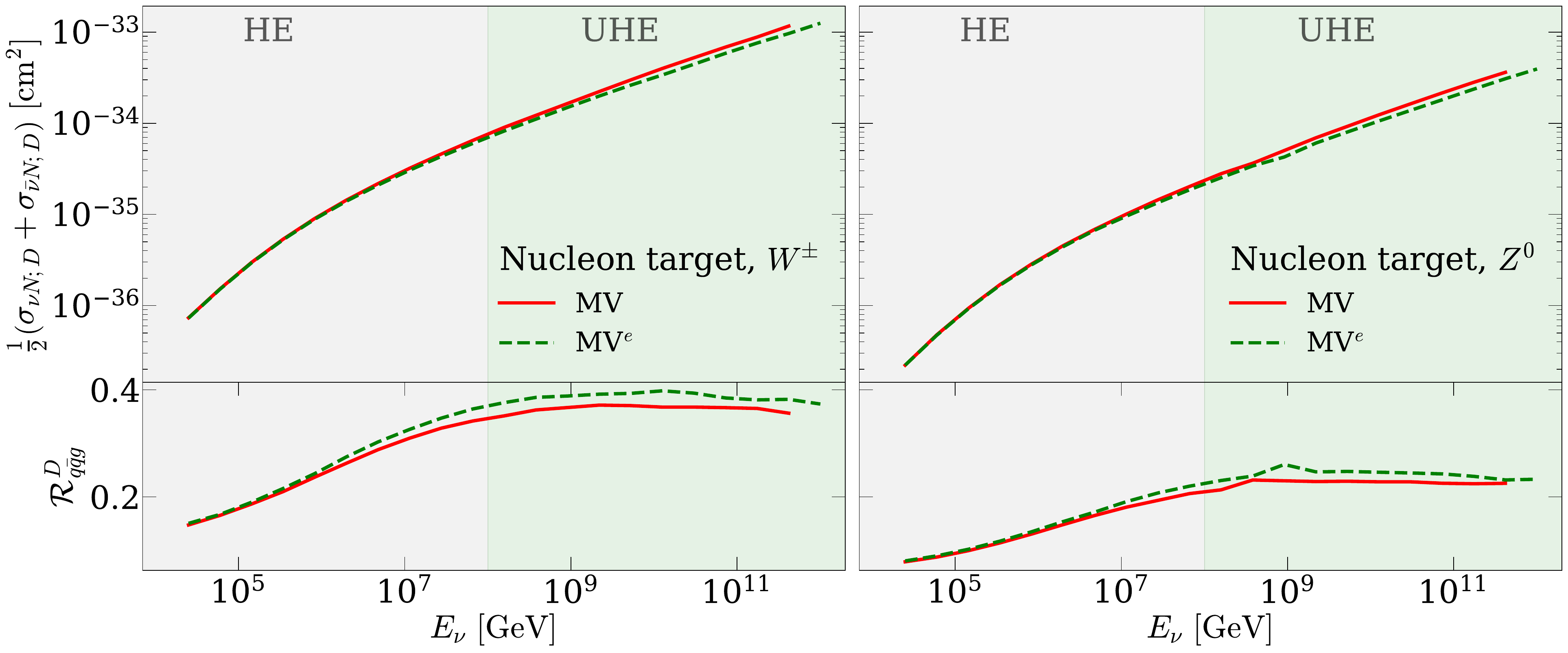}
    \caption{Diffractive neutrino-proton cross sections. The lower panels show the relative contribution from the $\ket{q\bar q g}$ Fock state. }
    \label{fig:dfrac_proton}
\end{figure*}

Diffractive cross sections off a nucleon target are shown in \cref{fig:dfrac_proton}.
The cross sections now rise more rapidly as a function of neutrino energy than in the case of an inclusive scattering. This can be seen from the following power-law approximations valid at $E_\nu \gtrsim 10^7$ GeV, compared to fits to inclusive cross section shown in Eqs.~\eqref{eq:edep_param_proton_mve} and~\eqref{eq:edep_param_proton_mv}.
\begin{align}
 \sigma^{CC}_{\nu N;D ({\rm MV}^e)} \simeq \left(1.798\times 10^{-37} ~{\rm cm^2}\right) \left(\frac{E_\nu}{1~\gev}\right)^{0.323}, \\
    \sigma^{CC}_{\nu N;D ({\rm MV})} \simeq \left(1.433\times 10^{-37} ~{\rm cm^2}\right) \left(\frac{E_\nu}{1~\gev}\right)^{0.339}.   
\end{align}
Diffraction accounts for only a very small fraction ($\sim 1\%$) of the total neutrino-nucleon cross section. This is significantly smaller than the fraction of $\sim 10\%$ observed in electron-proton scattering at HERA~\cite{ZEUS:1993vio,H1:1994ahk}. This is because  the diffractive contribution becomes important close to the black disc limit, which at high scales $Q^2$ probed in charged current exchanges is reached only at $E_\nu \gtrsim 10^{18}~ \gev$~\cite{Jalilian-Marian:2003ghc}.

The fraction of the $|q\bar q g\rangle$ state contribution to the diffractive cross section is shown in the lower panel in \cref{fig:dfrac_proton}. This fraction is defined as
\begin{equation}
    \mathcal{R}^D_{q\bar q g} = \frac{\sigma_{\nu A(N);D(q\bar q g)}}{\sigma_{\nu A(N);D(q\bar q g)} + \sigma_{\nu A(N);D(q\bar q)}}.
\end{equation}
Here $(q\bar q)$ and $(q\bar q g)$ correspond to the gauge boson Fock state included in the tree-level calculation.
The gluon component is found to have a significant contribution to the diffractive cross section, reaching up to 
$\sim 40\%$ for the charged-current exchange and $\sim 22\%$ for the neutral-current interaction at highest neutrino energies considered. In the HE region the relative contribution increases with increasing neutrino energy unitl a plateau is reached in the UHE regime.  Eventually this ratio is expected to decrease at very high neutrino energies.  This energy dependence of $\mathcal{R}^D_{q\bar q g}$  can be understood as follows. When the neutrino energy increases, the available phase space for the $\ket{q\bar qg}$ contribution, that typically correspond to larger invariant masses, increases. On the other hand, close to the black-disk limit (at very high energies)  the $\ket{q\bar q}$ contribution is maximally enhanced and the $\ket{q\bar qg}$ component vanishes~\cite{Kowalski:2008sa}. These two effects result in a plateau observed in the UHE regime . 

\begin{figure*}[ht!]
    \centering
    \includegraphics[width=\textwidth]{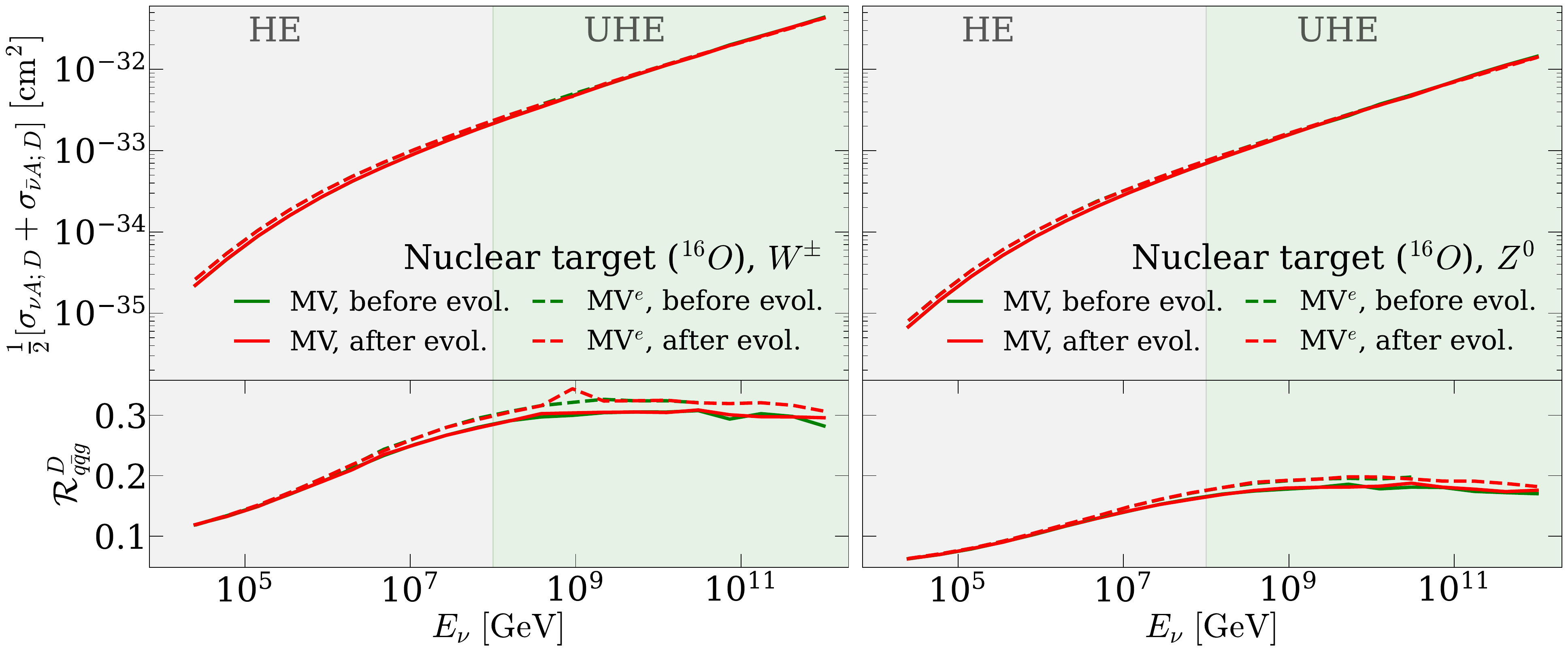}
    \caption{Diffractive neutrino-oxygen cross sections calculated using the dipole-proton amplitude generalized to the dipole-nucleus case using the two different prescriptions. The lower panels show the relative contribution from the $\ket{q\bar q g}$ Fock state.}
    \label{fig:dfrac_nucl_coh}
\end{figure*}

The coherent diffractive cross section for the neutrino-oxygen scattering is shown in \cref{fig:dfrac_nucl_coh}. Here we again show the dependence on the applied DIS fit and on the scheme used to generalize the dipole-nucleon amplitude to the dipole-nucleus case. Dependency on these choices is again weak. 
The relative importance of the $\ket{q\bar q g}$ Fock state is shown in the lower panel, and that is found to be slightly smaller than in the neutrino-proton scattering across all neutrino energies. This is because the nucleus is closer to the black-disc limit than the proton, and in that limit the $q\bar q$ component dominates.

\begin{figure*}[ht!]
    \centering
    \includegraphics[width=\textwidth]{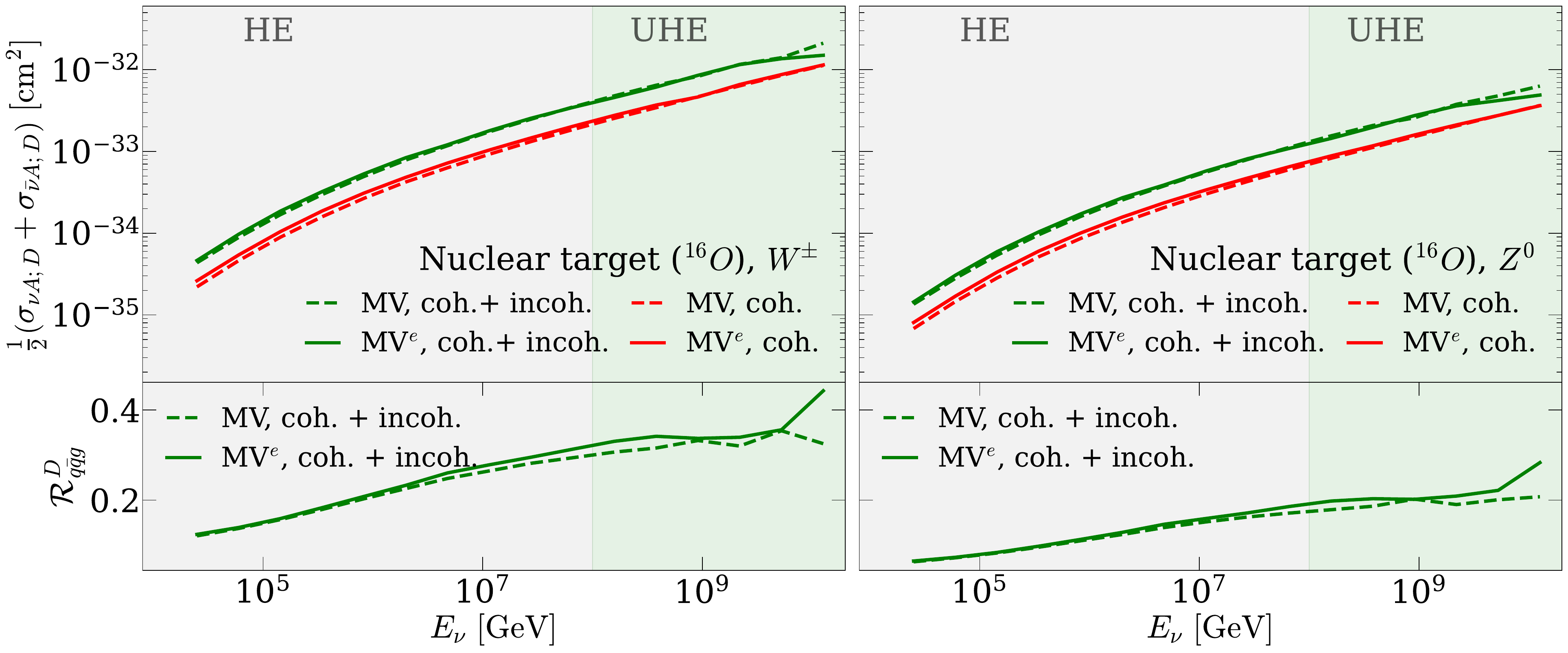}
    \caption{Diffractive neutrino-oxygen cross sections with (green) and without (red) incoherent contribution. The relative importance of the $\ket{q\bar q g}$ contribution is shown in the lower panel.
    }
    \label{fig:dfrac_nucl_incoh}
\end{figure*}

Next we include event-by-event fluctuations by sampling nucleon positions  from the Woods-Saxon distribution. These fluctuations have a negligible effect on the coherent cross section that is only sensitive to the average interaction, but result in a non-vanishing incoherent cross section. As we do not perform an impact parameter dependent evolution in this work, we only consider the {\em after evol.} setup. The obtained coherent and incoherent cross sections are shown in \cref{fig:dfrac_nucl_incoh}. 
The incoherent contribution is found to be numerically important, enhancing the diffractive cross section roughly by a factor of $2$.
Generically the incoherent cross section is expected to vanish in the black disc limit where fluctuations in the scattering amplitude disappear, but this behavior is not visible in the considered energy range.

The higher saturation scale of the nucleus relative to the proton should result in diffractive cross section being enhanced relative to the inclusive one. We find this fraction to be $2\%$ when the nucleus remains intact, and about $4\%$ when the target breakup is also included for the UHE neutrinos. This should be compared to $1\%$ obtained above for the neutrino-nucleon scattering. 
This means that while the nonlinear evolution has certain visible effects in the diffractive cross sections at highest considered neutrino energies $E_\nu \sim 10^{12}~\gev$ (e.g. $\sim 10\%$ nuclear suppression shown in Fig.~\ref{fig:inclusive_nucl}), in this energy domain the nucleus is still far from the black-disc limit where diffraction becomes important, and asymptotically a $50\%$ contribution to the total cross section.

The relative contribution from the $\ket{q\bar q g}$ state in the neutral-current exchange is approximately half of that in the charged-current exchange. This is due to the helicity structure which suppresses the gluon emission in the case of a $Z$ boson exchange by a factor of $2$, compared to the $W^\pm$ case. In particular, the structure function ratios (neglecting the difference in the $W^\pm$ and $Z$ masses) become
\begin{subequations}
    \begin{equation}
    \frac{F^{D(3);W^{\pm}}_{T/L;q\bar q}}{F^{D(3);Z}_{T/L;q\bar q}} = \frac{\sum\limits_{\rm ff'}{|V_{\rm ff'}|^2}}{\frac{1}{2}\sum\limits_{f} [(c_v^f)^2 + (c_a^f)^2]},
\end{equation} 
and
\begin{equation}
    \frac{F^{D(3);W^{\pm}}_{T;q\bar qg}}{F^{D(3);Z}_{T;q\bar qg}} = \frac{\sum\limits_{\rm ff'}{|V_{\rm ff'}|^2}}{\frac{1}{4}\sum\limits_{f} [(c_v^f)^2 + (c_a^f)^2]},
\end{equation} 
\end{subequations}
see \cref{app:coheren_diffraction,app:incoherent_diffraction}.    

\section{Summary and outlook}
\label{sec:conclusions}

We have calculated inclusive and diffractive neutrino-proton and neutrino-nucleus cross sections in the dipole picture at very high neutrino energies $E_\nu \gtrsim 10^5\,\gev$. The energy dependence is predicted by solving the non-linear Balitsky-Kovchegov evolution equation with the initial condition constrained by electron-proton DIS data from HERA. The applied setup is found to describe the available neutrino-nucleon total cross section data well, especially when the MV$^e$ fit from Ref.~\cite{Lappi:2013zma} is used. 

The dipole-nucleus scattering amplitude is obtained from the dipole-nucleon one using the optical Glauber model for a finite-$A$ nuclei. This ``nucleation'' can be implemented either before or after the BK evolution, which defines the two different prescriptions referred to as \emph{before evol.} and \emph{after evol.}, respectively. Comparing predictions obtained with both of these setups allows us to quantify the importance of saturation effects in the small-$x$ evolution in the case of neutrino-DIS. 

The inclusive neutrino-oxygen cross section is found to be sensitive to saturation effects in the UHE regime where the neutrino energy exceeds $10^8\,\gev$. In particular, a $\sim 10\%$ nuclear suppression is obtained in the (more realistic) \emph{before evol.} setup in this energy domain due to the non-linearities in the small-$x$ evolution of the nuclear target. 

When calculating diffractive cross sections we include 
the dominant high-$Q^2$ contribution from the $\ket{q\bar q g}$ Fock state of the heavy gauge boson. This contribution is found to be numerically significant, corresponding to $\sim 10\dots 40\%$ of the diffractive cross section in the considered energy range. 
Similarly incoherent diffraction, calculated by including event-by-event fluctuating nucleon positions, is estimated to account for a large fraction $\sim 50\%$ of the diffractive neutrino-nucleus cross section. 

The diffractive cross section accounts for a small fraction of the total cross section in the considered energy range ($10^4 < E_\nu\lesssim 10^{12}$~GeV).
It is about $1\%$ for a nucleon and $4\%$ for the oxygen, which demonstrates that even at the highest considered energies the nucleus is still far from the black disc limit where this ratio becomes $50\%$.

Our findings demonstrate that even with light nuclear targets such as oxygen far from the black disc regime, neutrino-nucleus scattering in the UHE kinematics can be sensitive to saturation effects.
While nuclear effects have also been considered in previous studies, such as in Ref.~\cite{Bhattacharya:2016jce} in terms of nuclear parton distribution functions (nDPFs), our calculation does not rely on a parametrized $x$ and $A$ dependence like nPDFs. Instead, nuclear modifications are obtained directly as a result of non-linear QCD dynamics. 
In particular saturation effects can be expected to be visible at the future measurements performed at IceCube-2~\cite{IceCube-Gen2:2020qha}. 

\begin{acknowledgments}
We thank Tuomas Lappi and Spencer Klein for fruitful discussions.
This work was supported by the Research Council of Finland, the Centre of Excellence in Quark Matter (project 346324 and 364191) and projects 338263, 346567 and 359902, and by the European Research Council (ERC, grant agreements ERC-2023-101123801 GlueSatLight and ERC-2018-ADG-835105 YoctoLHC). 
Computing resources from CSC – IT Center for Science in Espoo, Finland were used in this work.
The content of this article does not reflect the official opinion of the European Union and responsibility for the information and views expressed therein lies entirely with the authors. 

\end{acknowledgments}

\begin{appendix}
\section{Light-cone wave functions (LCWF) of vector bosons in the limit of massless quarks}
\label{app:wavefunctions}

\subsection{Coordinate-space tree-level $\ket{q\bar q}$ LCWF}

The squared wave functions for the tree-level $\ket{q\bar{q}}$ component of the massive vector bosons' quantum state can be found in Ref.~\cite{Barone:1992aw} for massive (anti)quarks and Refs.~\cite{Kutak:2003bd,Albacete:2015zra} for massless (anti)quarks. Here we quote the expressions for the overlap functions in the massless limit, which are defined by
\begin{equation}
    \Phi^{V}_{\lambda} = \sum_{h_0,h_1} [\Psi^{V\to q\bar q}_{\lambda;h_0,h_1}(\rt',z,Q^2)]^{*} \Psi^{V\to q\bar q}_{\lambda;h_0,h_1}(\rt,z,Q^2),
\end{equation}
where $\Psi^{V\to q\bar q}_{\lambda;h_0,h_1}(\rt,z,Q^2)$ is the wave function for the vector boson $V$ of polarization $\lambda$ and virtuality $Q^2$ to fluctuate into a dipole of transverse size $\rt$ consisting a quark and an antiquark of helicities $h_0$ and $h_1$, respectively. Here $z$ is the fraction of the boson  longitudinal momentum carried by the quark. The squared wave functions can be obtained simply by putting $\rt' = \rt$.

 The overlap functions for the $W^\pm$ boson with different polarization states read    
    \begin{subequations}
    \label{seq:W_overlap}
        \begin{equation}
        \begin{aligned}
            &\Phi_{\lambda=0}^{W^\pm} (\rt,\rt',z, Q^2) = \frac{4N_c (4\pi\alpha_W)}{\pi^2}\sum\limits_{\rm ff'} |V_{\rm ff'}|^2 \\
            &\quad\qquad \times Q^2 z^3(1-z)^3 K_0(\bQ r)K_0(\bQ r'),
        \end{aligned}
        \end{equation}
        \begin{equation}
        \begin{aligned}
            &\Phi_{\lambda=1}^{W^\pm} (\rt, \rt', z, Q^2)= \frac{2N_c (4\pi\alpha_W)}{\pi^2} \sum\limits_{\rm ff'}|V_{\rm ff'}|^2 \\
            & \qquad \times z(1-z)^3\bQ^2 K_1(\bQ r)K_1(\bQ r') \frac{\rt\cdot\rt'}{rr'}, \ \text{and}
        \end{aligned} 
        \end{equation}
        \begin{equation}
        \begin{aligned}
            &\Phi_{\lambda=-1}^{W^\pm} (\rt,\rt', z, Q^2)= \frac{2N_c (4\pi\alpha_W)}{\pi^2} \sum\limits_{\rm ff'}|V_{\rm ff'}|^2 \\ & \qquad \times z^3(1-z)\bQ^2 K_1(\bQ r)K_1(\bQ r') \frac{\rt\cdot\rt'}{rr'}.
        \end{aligned} 
        \end{equation}
    \end{subequations}
    Similarly for the $Z$ boson we obtain
    \begin{subequations}
    \label{seq:Z_overlap}
        \begin{equation}
        \begin{aligned}
             &\Phi_{\lambda=0}^{Z} (\rt, \rt', z, Q^2) =   \sum_f\left[(c_v^f)^2 + (c_a^f)^2 \right] \\
             &\quad\times \frac{2N_c (4\pi\alpha_Z)}{\pi^2} Q^2z^3(1-z)^3 K_0(\bQ r)K_0(\bQ r'),
        \end{aligned}
        \end{equation}
        \begin{equation}
        \begin{aligned}
            &\Phi_{\lambda=1}^{Z} (\rt,\rt',z,Q^2) =  z(1-z) \bQ^2 K_1(\bQ r)K_1(\bQ r') \frac{\rt\cdot\rt'}{rr'} \\
            & \quad\times \frac{N_c (4\pi\alpha_Z)}{2\pi^2} \sum_f\left[z^2(c_v^f-c_a^f)^2 + (1-z)^2(c_v^f+c_a^f)^2\right],       
        \end{aligned}
        \end{equation}
        \begin{equation}
        \begin{aligned}
           & \Phi_{\lambda=-1}^{Z} (\rt,\rt',z,Q^2) =   z(1-z) \bQ^2 K_1(\bQ r)K_1(\bQ r') \frac{\rt\cdot\rt'}{rr'} \\
           & \quad \times\frac{N_c (4\pi\alpha_Z)}{2\pi^2}\sum_f\left[z^2(c_v^f+c_a^f)^2 + (1-z)^2(c_v^f-c_a^f)^2\right].
        \end{aligned}
        \end{equation}
    \end{subequations}
In the case of transverse polarization $\lambda=\pm 1$ there are also asymmetric terms proportional to the Levi-Civita symbol $\epsilon^{ij}$ in the wave function overlaps. These terms are not included in the above expressions as as they eventually give zero contributions to the cross sections. In the above expressions, $\nc = 3$ is the number of colors, and $\bQ^2 = Q^2z(1-z)$. Each complex number $V_{\rm ff'}$ is the CKM matrix element for the transition between the flavors $f$ and $f'$ by charged-current exchange. The real numbers $c_{v,a}^f$ are the vector ($v$) and axial-vector ($a$) coefficients for the corresponding neutral-current splitting $Z^0\to f\Bar{f}$. Their values  read
\begin{equation}
    \begin{aligned}
        c_{v}^{u,c,t} &= \frac{1}{2}-\frac{4}{3}\sin^2\theta_W,~ c_a^{u,c,t} = \frac{1}{2},\\
        c_{v}^{d,s,b} &= -\frac{1}{2}+\frac{2}{3}\sin^2\theta_W,~ c_a^{d,s,b} = -\frac{1}{2},
    \end{aligned}
\end{equation}
 where $\theta_W$ is the Weinberg angle. The summations are performed over active flavours, which is light quark flavours in our calculations consistently with the applied DIS fit~\cite{Lappi:2013zma}. The couplings $\alpha_{W,Z}$ are given by
\begin{subequations}
\begin{equation}
    \alpha_W = \frac{e^2}{32\pi \sin^2\theta_W},
\end{equation}
and
\begin{equation}
    \alpha_Z = \frac{e^2}{16\pi \sin^2\theta_W\cos^2 \theta_W}.
\end{equation}    
\end{subequations}

\subsection{Momentum-space tree-level $\ket{q\bar q g}$ LCWF} 
\label{subapp:qqg_wf}
We follows Refs.~\cite{Beuf:2017bpd,Iancu:2022lcw} to derive the tree-level $\ket{q(k_0)\bar q (k_1) g (k_2)}$ LCWF of the transverse vector boson with four-momentum $q = \left(q^{+},-\frac{Q^2}{2{q+}},0_{\perp}\right)$ in the large-$Q^2$ limit. The general expression is given by
\begin{widetext}
    \begin{equation}
    \label{eq:qqg_wf_gen}
    \begin{aligned}
        &\Psi^{W^{\pm}/Z\to q\bar qg}_{\lambda=\pm 1} = \frac{(2\pi)^3\delta^{3}\left(\vk_0 + \vk_1 + \vk_2 - \vq \right)}{\mathrm{ED}_{012}} A^{W/Z}g(t^{a_2})_{\alpha_0\alpha_1} \\
        & \times \left\{ \int \frac{\dd^3\vk_{0'}}{(2\pi)^3} \frac{\Theta(k_{0'}^{+})}{2k_{0'}^{+}} \frac{(2\pi)^3\delta^3\left(\vk_{0} + \vk_{2} - \vk_{0'}\right)}{\mathrm{ED}_{0'1}} \sum_{h_{0'}} \left[\bar{u}(0) \gamma^{\mu} \epsilon^{*}_{\mu}(k_2;\lambda_2) u(0') \right]\left[\bar{u}(0') \gamma^{\nu} (c_v^f-c_a^f \gamma^5)\varepsilon_\nu(q;\lambda) v(1) \right] \right. \\
        & \left. - \int \frac{\dd^3\vk_{1'}}{(2\pi)^3} \frac{\Theta(k_{1'}^{+})}{2k_{1'}^{+}} \frac{(2\pi)^3\delta^3\left(\vk_{1} + \vk_{2} - \vk_{1'}\right)}{\mathrm{ED}_{01'}} \sum_{h_{1'}} \left[\bar{u}(0) \gamma^{\mu} (c_v^f-c_a^f \gamma^5)\varepsilon_{\mu}(q) v(1') \right]\left[\bar{v}(1') \gamma^{\nu} \epsilon^{*}_{\nu}(k_2;\lambda_2) v(1) \right] \right. \\
        & \left. +  \frac{1}{2(k_0^{+} + k_2^{+})} \left[ \bar{u}(0) \gamma^{\mu} \epsilon^{*}_{\mu}(k_2;\lambda_2) \gamma^{+} \gamma^{\nu} (c_v^f-c_a^f \gamma^5)\varepsilon_\nu(q) v(1)  \right] \right. \\
        & \left. - \frac{1}{2(k_1^{+} + k_2^{+})} \left[ \bar{u}(0) \gamma^{\mu} (c_v^f-c_a^f \gamma^5)\varepsilon_{\mu}(q;\lambda) \gamma^+ \gamma^{\nu} \epsilon^{*}_{\nu}(k_2;\lambda_2) v(1) \right] \right\}
    \end{aligned}
    \end{equation}
\end{widetext}
where $\vk=(k^+,\kt)$, and $\alpha_{0,1}$ and $a_2$ are the color indices of (anti)quark and gluon, respectively, and $u(i) \equiv u_{h_i}(k_i)$ and $v(i) \equiv v_{h_i}(k_i)$ are Dirac spinors for quark and antiquark of helicity $h_i$. The vector and axial-vector coefficients for the charged-current exchange are identically unity, $c_v^f = c_a^f = 1$. The prefactors $A^{W/Z}$ read
\begin{equation}
    A^{W} = \frac{eV_{\rm ff'}}{2\sqrt{2}\sin\theta_W},~{\rm and~}A^{Z} = \frac{e}{2\sin\theta_W\cos\theta_W}. 
\end{equation}
The transverse polarization vectors for the vector bosons, $\varepsilon_{\mu}(q;\lambda)$, and for the (on-shell) gluon, $\epsilon_{\mu}(k_2;\lambda_2)$, are
\begin{subequations}
    \begin{equation}
        \varepsilon^{\mu}(q;\lambda=\pm 1) = (0,0,\varepsilon_{\perp}(\lambda)),
    \end{equation}
    \begin{equation}
        \epsilon^{\mu}(k_2;\lambda_2=\pm 1) = \left(0,\frac{\epsilon_{\perp}\cdot \kt_2}{k_2^+},\epsilon_{\perp}(\lambda_2)\right),
    \end{equation}
    %*
    \begin{equation}
        \varepsilon_{\perp}(\lambda) = \epsilon_{\perp}(\lambda) = -\frac{1}{\sqrt{2}}(\lambda,i).
    \end{equation}
\end{subequations}
Additionally, the energy denominators $\mathrm{ED}_{012}, \mathrm{ED}_{0'1}$ and $\mathrm{ED}_{01'}$ read
%*
\begin{subequations}
    \begin{equation}
    \begin{aligned}
    \mathrm{ED}_{012} &\equiv - \frac{Q^2}{2q^+} - \frac{\kt_0^2}{2k_0^+} - \frac{\kt_1^2}{2k_1^+} - \frac{\kt_2^2}{2k_2^+} \\
    &= -\frac{1}{2q^{+}}\left(Q^2 + \frac{\kt_0^2}{z_0} + \frac{\kt_1^2}{z_1} + \frac{\kt_2^2}{z_2}\right) ,
    \end{aligned}
    \end{equation}    
    \begin{equation}
        \mathrm{ED}_{0'1} \equiv -\frac{1}{2z_{0'}z_1q^+} \left(z_{0'}z_1Q^2 + \kt_1^2\right),
    \end{equation}
    \begin{equation}
        \mathrm{ED}_{01'} \equiv -\frac{1}{2z_{0}z_{1'}q^+} \left(z_{0}z_{1'}Q^2 + \kt_0^2\right),
    \end{equation}
\end{subequations}
where $z_{i} \equiv k_i^+/q^{+}$ is the longitudinal momentum fraction of the corresponding parton of momentum $k_i$, and the energy-momentum conservation at vertices are implicitly implemented. The first and the second term in the curly bracket of \cref{eq:qqg_wf_gen} correspond to a causal emissions of gluon from the quark or from the antiquark, respectively. The last two terms corresponds to instantaneous gluon emissions. It is convenient to use the following transverse momentum variables instead of the momenta $\kt_{0,1,2}$:
\begin{equation}
    \begin{aligned}
        \Pt_{q\bar q} & = \frac{z_1\kt_0-z_0\kt_1}{z_0+z_1},\\
        \Kt_{gg} & = (z_0+z_1)\kt_2 - z_2(\kt_0 + \kt_1),
    \end{aligned}
\end{equation}
which can be interpreted as the relative momenta between the quark and the antiquark, and between the gluon and the quark-antiquark dipole, respectively. In the large-$Q^2$ limit, we have a strong ordering in the longitudinal momenta, either $z_2\ll z_1 \ll z_0$ or $z_2\ll z_0 \ll z_1$. These two configurations give the same contributions, and hence, we only consider the former one from now on. In such so-called aligned-jet configurations, we have the following hierachy in the transverse momenta: $\Kt_{gg} \ll \Pt_{q\bar q} \sim Q $. It means that, in the transverse coordinate space, the gluon is emitted far away from the quark-antiquark dipole, and hence, this parent dipole can be seen as an effective gluon. The three-particle system then effectively becomes a gluon dipole $gg$. Performing algebras on the Dirac matrix elements in this approximate kinematics using basic e.g. from Refs.~\cite{Angelopoulou:2023qdm,Kovchegov:2012mbw} and keeping the terms up to the next-to-eikonal accuracy in the energy denominators, we eventually get
\begin{subequations}
\label{seq:qqg_wf}
\begin{equation}
    \begin{aligned}
        &\Psi_{\lambda=\pm 1}^{W^{\pm}\to q\bar qg} \underset{large-Q^2}{\simeq} A^{W} g(t^{a_2})_{\alpha_0\alpha_1}\sqrt{z_0z_1} \\
        &\times \frac{2\delta_{h_0,-h_1}(1-2h_0)(1-\lambda)\epsilon^{j}_{\perp}(\lambda)\epsilon^{*l}_{\perp}(\lambda_2)}{z_2\left(Q^2 + \frac{\Pt^2_{q\bar q}}{z_1} + \frac{\Kt^2_{gg}}{z_2}\right)} \\
        &\times \left[-\frac{z_2}{z_1}\frac{\Pt^{j}_{q\bar q} \Pt^{l}_{q\bar q}}{z_1 Q^2 + \Pt^2_{q\bar q}} + \frac{\Kt^{j}_{gg} \Kt^{l}_{gg}}{z_1 Q^2 + \Pt^2_{q\bar q}} \right.\\
        &\left.-\frac{2\Pt^{j}_{q\bar q} \Pt^{i}_{q\bar q}\Kt^{i}_{gg} \Kt^{l}_{gg}}{(z_1 Q^2 + \Pt^2_{q\bar q})^2} + \frac{z_2}{2z_1}\delta^{jl}\right],
    \end{aligned}
\end{equation}
\begin{equation}
    \begin{aligned}
        &\Psi_{\lambda=\pm 1}^{Z\to q\bar qg} \underset{large-Q^2}{\simeq} A^{Z}(t^{a_2})_{\alpha_0\alpha_1}\sqrt{z_0z_1} \\
        &\times \frac{4\delta_{h_0,-h_1}\delta_{2h_0,\lambda}(c_v^f- \lambda c_a^f)\epsilon^{j}_{\perp}(\lambda)\epsilon^{*l}_{\perp}(\lambda_2)}{z_2\left(Q^2 + \frac{\Pt^2_{q\bar q}}{z_1} + \frac{\Kt^2_{gg}}{z_2}\right)} \\
        &\times \left[-\frac{z_2}{z_1}\frac{\Pt^{j}_{q\bar q} \Pt^{l}_{q\bar q}}{z_1 Q^2 + \Pt^2_{q\bar q}} + \frac{\Kt^{j}_{gg} \Kt^{l}_{gg}}{z_1 Q^2 + \Pt^2_{q\bar q}} \right.\\
        &\left.-\frac{2\Pt^{j}_{q\bar q} \Pt^{i}_{q\bar q}\Kt^{i}_{gg} \Kt^{l}_{gg}}{(z_1 Q^2 + \Pt^2_{q\bar q})^2} + \frac{z_2}{2z_1}\delta^{jl}\right],
    \end{aligned}
\end{equation}
\end{subequations}
We find that the wave functions of both $W^{\pm}$ and $Z$ have the same kinematical structure as is obtained for virtual photon in Ref.~\cite{Beuf:2022kyp}. The main difference between them lies in the helicity structure. In particular, in the charged-current interaction, the helicity of the (anti)quark is fixed, which is due to the fact that, in the chiral limit, the $W^{\pm}$ bosons couple only to left-handed quarks and right-handed antiquarks. The polarization state of $W^{\pm}$ is also selected in each aligned-jet configuration: $\lambda=-1$ when the quark leg carries the largest momentum fraction, and $\lambda=1$ in the remaining case. Those are not realized in the case of the neutral-current exchange, as the Dirac matrix element for the $Z\to q\bar q$ vertex mixes the left-handed and the right-handed projections.   

The obtained wave functions in \cref{seq:qqg_wf} are written in terms of kinematic variables before the interaction with the target. To facilitate the derivation of diffractive structure functions (see \cref{app:coheren_diffraction,app:incoherent_diffraction}), let us turn on the interaction by exchanging a pomeron state. In the infinite momentum frame where the target has a large longitudinal (minus) momentum, the invariant $\xpom$ can be interpreted as the fraction of the target's longitudinal momentum carried by the pomeron. The invariant $\beta\equiv\xbj/\xpom$ is, in this frame, the fraction of pomeron's minus-component momentum carried by the struck parton. We can define $z$ to be the fraction of pomeron's minus-component momentum transferred to the quark-antiquark system, and define also $\xi \equiv \beta/z$. The use of these variables allows to make the connection to the Wusthoff's result in the case of a virtual photon exchange~\cite{Wusthoff:1997fz,Golec-Biernat:1999qor,Kowalski:2008sa}. Due to the presence of the interaction, let us define $\hat{\Kt}_{gg}$ and $\hat{\Pt}_{q\bar q}$ to be the analog variables to $\Kt_{gg}$ and $\Pt_{q\bar q}$, but after the interaction. At large-$Q^2$, since the $q\bar q$ dipole effectively shrinks to a point-like particle, $\hat{\Pt}_{q\bar q} = \Pt_{q\bar q}$. We have the following relations~\cite{Beuf:2022kyp}:
\begin{align}
    z_2 &= \frac{\beta}{1-z} \frac{\hat{\Kt}_{gg}}{Q^2},\\
    \Pt_{q\bar q}^2 + z_1Q^2 &= \frac{z_1Q^2}{\xi},\\
    Q^2 + \frac{\Pt_{q\bar q}^2}{z_1} + \frac{\Kt_{gg}^2}{z_2} &= \frac{Q^2}{\beta\hat{\Kt}_{gg}^2} \left[z\hat{\Kt}_{gg}^2 + (1-z)\Kt_{gg}^2\right].
\end{align}
%s
Using these new variables and performing possible angular averages, we obtain the following momentum-space wave-function overlap for the $W^{\pm}$ bosons:
\begin{equation}
    \begin{aligned}
        &\Tilde{\Phi}_T^{W^{\pm}} \equiv \frac{1}{2} \sum_{\rm f,f'}\sum_{\{h_{0,1},\lambda_2,\lambda\}}\sum_{\{\alpha_{0,1},a_2\}}\left(\Psi_{\lambda=\pm 1}^{W^{\pm}\to q\bar q g}\right)^{\dagger}\Psi_{\lambda=\pm 1}^{W^{\pm}\to q\bar q g} \\
        = & \sum\limits_{\rm f,f'}|V_{\rm ff'}|^2\frac{4e^2g^2 C_F N_C}{\sin^2\theta_W} \beta^2 \frac{(1-z)^2}{z^2} \frac{z_0}{z_1Q^4}\left[\xi^2 + (1-\xi)^2\right] \\
        \times & \frac{\Kt_{gg}^j \Kt_{gg}^l + \frac{z}{2(1-z)}\hat{\Kt}_{gg}\delta^{jl}}{\left[z\hat{\Kt}_{gg}^2 + (1-z)\Kt_{gg}^2\right]}~  \frac{{\Kt'}_{gg}^j {\Kt'}_{gg}^l + \frac{z}{2(1-z)}\hat{\Kt}_{gg}\delta^{jl}}{\left[z\hat{\Kt}_{gg}^2 + (1-z){\Kt'}^{2}_{gg}\right]},
    \end{aligned}
\end{equation}
where the prime notation is for the complex conjugated wave function. 

Repeating the same procedure for the neutral-current exchange, we get
\begin{equation}
    \begin{aligned}
        &\Tilde{\Phi}_T^{Z} \equiv \frac{1}{2} \sum_{f}\sum_{\{h_{0,1},\lambda_2,\lambda\}}\sum_{\{\alpha_{0,1},a_2\}}\left(\Psi_{\lambda=\pm 1}^{Z\to q\bar q g}\right)^{\dagger}\Psi_{\lambda=\pm 1}^{Z\to q\bar q g} \\
        = & \sum\limits_{f}\left[(c_v^f)^2+(c_a^f)^2\right]\frac{2e^2g^2 C_F N_C}{\sin^2\theta_W \cos^2\theta_W}  \\
        &\times \beta^2 \frac{(1-z)^2}{z^2} \frac{z_0}{z_1Q^4} \left[\xi^2 + (1-\xi)^2\right] \\
        \times & \frac{\Kt_{gg}^j \Kt_{gg}^l + \frac{z}{2(1-z)}\hat{\Kt}_{gg}\delta^{jl}}{\left[z\hat{\Kt}_{gg}^2 + (1-z)\Kt_{gg}^2\right]}~  \frac{{\Kt'}_{gg}^j {\Kt'}_{gg}^l + \frac{z}{2(1-z)}\hat{\Kt}_{gg}\delta^{jl}}{\left[z\hat{\Kt}_{gg}^2 + (1-z){\Kt'}^{2}_{gg}\right]}.
    \end{aligned}
\end{equation}

\section{Tree-level $q\bar q$ and $q\bar q g$ contributions to coherently diffractive structure functions}
\label{app:coheren_diffraction}

The (coherently) diffractive structure functions for the exchange of a virtual photon can be found e.g. from Ref.~\cite{Kowalski:2008sa}. In this appendix, we reported the results for the (coherently) diffractive structure functions for the exchange of massive vector bosons, which are used to make the first estimations for the diffractive cross-sections presented in the main text.   

\subsection{The $\ket{q\bar q}$ contributions}
Following Ref.~\cite{Beuf:2022kyp}, the contributions from the $\ket{q\bar q}$ Fock component to the diffractive structure functions can be computed as
\begin{widetext}
\begin{subequations}
\label{eq:coh_qq}
\begin{equation}
\label{eq:formul_FL_qq}
\begin{aligned}
    \xpom F^{D(3);W^{\pm}/Z}_{q\bar q;L} =& \frac{1}{4\pi\alpha_{W/Z}}\frac{ Q^4 \xpom}{4\pi^2\xbj} \int_0^{\infty} \dd |t| \int_0^1 \dd z  \int \dd^2\rt \int\dd^2 \rt' \int\dd^2\bt\int\dd^2\bt' \Theta(\kappa^2) \mathcal{I}_{t}^{(2)} \mathcal{I}_{\kappa}^{(2)} \\
    &\times \Phi_{\lambda=0}^{W^{\pm}/Z}(\rt,\rt',z,Q^2)N(\rt,\bt,\xpom)N(\rt',\bt',\xpom),
\end{aligned}
\end{equation}
and
\begin{equation}
\label{eq:formul_FT_qq}
\begin{aligned}
    \xpom F^{D(3);W^{\pm}/Z}_{q\bar q;T} =& \frac{1}{4\pi\alpha_{W/Z}}\frac{ Q^4 \xpom}{4\pi^2\xbj} \int_0^{\infty} \dd |t| \int_0^1 \dd z \int \dd^2\rt \int\dd^2 \rt' \int\dd^2\bt\int\dd^2\bt' \Theta(\kappa^2) \mathcal{I}_{t}^{(2)} \mathcal{I}_{\kappa}^{(2)} \\
    &\times\frac{1}{2}\sum_{\lambda=\pm 1} \Phi_{\lambda}^{W^{\pm}/Z}(\rt,\rt',z,Q^2) N(\xpom,\rt,\bt)N(\xpom, \rt',\bt'),
\end{aligned}
\end{equation}
\end{subequations}    
\end{widetext}
where $\kappa^2 = M_X^2 z(1-z) =\bQ^2\left[(\xpom/\xbj)-1\right] z(1-z)$ and the impact parameter is defined as $\bt=\frac{1}{2}\left( {\bf  x}_q + {\bf  x}_{\bar q} \right)$, with ${\bf  x}_{q(\bar q)}$ being the position of the (anti)quark leg. In \cref{eq:formul_FL_qq,eq:formul_FT_qq}, the functions $\mathcal{I}_{t}^{(2)}$ and $\mathcal{I}_{\kappa}^{(2)}$ are defined as
\begin{subequations}
    \begin{equation}
    \label{eq:It}
        \mathcal{I}_{t}^{(2)} = \int \frac{\dd^2\Delta}{(2\pi)^2} \delta\left(\Delta^2-|t|\right)e^{-i\Delta\cdot(\bt-\bt')},
    \end{equation}
    \begin{equation}
    \label{eq:Ik}
        \mathcal{I}_{\kappa}^{(2)} = \int \frac{\dd^2\lt}{(2\pi)^2} \delta\left(\lt^2-\kappa^2\right)e^{-i\lt\cdot(\rt-\rt')}.
    \end{equation}
\end{subequations}
In fact, an off-forward phase has been neglected in \cref{eq:It}, which is a reasonable approximation at large $Q^2$ or large $M_X^2$ (see Ref.~\cite{Beuf:2022kyp} for more details). We assume that the dipole-target scattering amplitude does not depend on the orientation of the dipole and of the impact parameter. Substituting the formulae \eqref{seq:W_overlap} and \eqref{seq:Z_overlap} for the wave function overlaps into \cref{eq:formul_FL_qq,eq:formul_FT_qq} and performing the integrals, we eventually get

\begin{subequations}
\label{seq:F_qq}
    \begin{equation}
        \begin{aligned}
        & \xpom F^{D(3);W^{\pm}}_{q\bar q;L} = \frac{N_cQ^6}{\pi^2 \beta} \sum_{\rm \rm ff'} |V_{\rm \rm ff'}|^2 \int\limits_0^1 \dd{z} z^3(1-z)^3 \Xi_0,
        \end{aligned}
    \end{equation}
    \begin{equation}
        \begin{aligned}
        \xpom F^{D(3);W^{\pm}}_{q\bar q;T} &= \frac{N_cQ^4}{4\pi^2 \beta} \sum_{\rm ff'} |V_{\rm ff'}|^2 \\
        &\hspace{-2.8em}\times \int_{0}^{1} \dd{z}  z(1-z) \bQ^2\left[z^2 + (1-z)^2\right] \Xi_1,
        \end{aligned}    
    \end{equation}
    \begin{equation}
        \begin{aligned}
            \xpom F^{D(3);Z}_{q \bar q;L} &= \frac{N_cQ^6}{2\pi^2 \beta} \sum_{f} \left[(c_v^f)^2 + (c_a^f)^2\right] \\
            &\qquad\times \int_0^1 \dd z z^3(1-z)^3 \Xi_0,
        \end{aligned}
    \end{equation}
    and
    \begin{equation}
        \begin{aligned}
             \xpom F^{D(3);Z}_{q \bar q;T} &= \frac{N_cQ^4}{8\pi^2 \beta} \sum_{f} \left[(c_v^f)^2 + (c_a^f)^2\right] \\
             & \hspace{-2.8em}\times\int_{0}^{1} \dd{z}  z(1-z) \bQ^2\left[z^2 + (1-z)^2\right] \Xi_1,
        \end{aligned}
    \end{equation}
\end{subequations}
where we have defined the following auxiliary functions
\begin{equation}
    \Xi_n =  \int \dd[2]\bt \left( \int \dd r r J_n (\kappa r )K_n(\bQ r) N(\xpom,r,b) \right)^2.
\end{equation}
\subsection{The transverse $\ket{q\bar q g}$ contribution}
At large-$Q^2$  the transverse $\ket{q\bar q g}$ contribution dominates over the longitudinal one. In this limit, as discussed in \cref{subapp:qqg_wf}, the aligned-jet configurations play the dominatant role, and the $q\bar q g$ system behaves like a dipole in the adjoint representation. Again, from Ref.~\cite{Beuf:2022kyp}, the starting point to compute the $\ket{q\bar q g}_T$ contribution to the diffractive structure functions is the following expression:  
%
%\newpage
%
\begin{widetext}
    \begin{equation}
    \label{eq:formul_Fqqg}
    \begin{aligned}
        &\xpom F_{q\bar q g;T}^{D(3);W^{\pm}/Z} = \frac{1}{4\pi\alpha_{W/Z}}\frac{Q^4\xpom}{16\pi^3 \xbj} \int_{0}^{+\infty} \dd|t|\int_0^1 \frac{\dd z_0}{z_0} \int_0^1 \frac{\dd z_1}{z_1} \int_0^1 \frac{\dd z_2}{z_2} \delta(1- z_0 - z_1 - z_2) \\
        & \times \int \frac{\dd^2\hat{\Pt}_{q\bar q}}{(2\pi)^2} \int \frac{\dd^2\hat{\Kt}_{gg}}{(2\pi)^2} \int \frac{\dd^2\Delta}{(2\pi)^2} \delta\left(\Delta^2 - t\right) \delta\left(\frac{\hat{\Kt}_{gg}^2}{z_2(z_0+z_1)} + \frac{z_0 + z_1}{z_0z_1}\hat{\Pt}_{q\bar q} - M_X^2\right)\\
        &\times (2\pi)^6 \int \frac{\dd^2 \ut}{2\pi} \int \frac{\dd^2 \rt}{2\pi} \int \frac{\dd^2 \bt}{2\pi} \int \frac{\dd^2 \ut'}{2\pi} \int \frac{\dd^2 \rt'}{2\pi} \int \frac{\dd^2 \bt'}{2\pi} e^{-i(\ut - \ut')\cdot \hat{\Pt}_{q\bar q}} e^{-i(\rt-\rt')\cdot \hat{\Kt}_{gg}} e^{-i(\bt-\bt')\cdot \Delta} \\
        &\times \int \frac{\dd^2 \Pt_{q\bar q}}{(2\pi)^2} \int \frac{\dd^2 \Kt_{q\bar q}}{(2\pi)^2} \int \frac{\dd^2 \Pt'_{q\bar q}}{(2\pi)^2} \int \frac{\dd^2 \Kt'_{q\bar q}}{(2\pi)^2} e^{i\ut\cdot \Pt_{q\bar q} - i\ut'\cdot\Pt'_{q\bar q}}e^{i\rt\cdot \Kt_{gg} - i\rt'\cdot\Kt'_{gg}}~\Tilde{\Phi}^{W^{\pm}/Z }_T \Tilde{N}_{\ut\rt\bt}\Tilde{N}_{\ut'\rt'\bt'},
    \end{aligned}
    \end{equation}    
\end{widetext}
In the above equation, $\ut$ ($\ut'$), $\rt$ ($\rt'$) and $\bt$ ($\bt'$) are the $q\bar q$ dipole size, the effective size of the gluon dipole, and the center of mass of the three-body $q\bar q g$ system in the (complex-conjugated) amplitude, respectively. The scattering amplitude for the $q\bar qg$-target interaction is denoted by $\tilde{N}_{\ut\rt\bt}$. In the large-$Q^2$ limit the parent dipole is small, $|\ut| \ll |\rt|$, and effectively $\tilde{N}_{\ut\rt\bt}$ corresponds to a dipole-target scattering amplitude in the adjoint representation for the gluonic dipole of size $\rt$.

Using the wave-function overlap given in \cref{subapp:qqg_wf}, and performing the step-by-step simplification following the calculations in Ref.~\cite{Beuf:2022kyp}, we can obtain the Wusthoff-like expressions~\cite{Wusthoff:1997fz,Golec-Biernat:1999qor,Kowalski:2008sa} of the diffractive structure functions for the exchange of massive vector bosons. One important step in the calculation is to transform from $(z_2,\hat{\Kt}_{gg}^2)$ into $(z,k^2)$ given by
\begin{equation}
    1-z = \frac{\beta \hat{\Kt}_{gg}}{Q^2 z_2},~k^2 = \frac{\hat{\Kt}_{gg}}{1-z}.
\end{equation}
The interpretation of $z$ is given in \cref{subapp:qqg_wf}, while $k^2$ is the mean virtuality of the exchanged $t$-channel gluon in the two-gluon exchange mode~\cite{Wusthoff:1997fz,Golec-Biernat:1999qor}. In the end, we obtain
\begin{widetext}
\begin{subequations}
\label{eq:qqg_diff_structurefun}
    \begin{equation}
        \begin{aligned}
            \xpom F_{T,q\bar q g}^{D(3),W^\pm} 
             & =  \frac{\alpha_s(Q^2)\beta}{(2\pi)^3} \sum_{\rm ff'}|V_{\rm ff'}|^2 \int\limits_\beta^1 \dd{z} \left[\left(\frac{\beta}{z}\right)^2 + \left(1-\frac{\beta}{z}\right)^2\right] \int\limits_0^{Q^2} \dd{k^2} k^4 \ln \frac{Q^2}{k^2} \\
            & \qquad\times\int \dd{b} b \left\{ \int \dd{r} r J_2(\sqrt{1-z}kr)K_2(\sqrt{z}kr)[2\Tilde{N}(\xpom,\rt;\bt)]\right\}^2,
        \end{aligned}
    \end{equation}
    \begin{equation}
        \begin{aligned}
       \xpom F_{T,q\bar q g}^{D(3),Z} & =  \frac{\alpha_s(Q^2)\beta}{4(2\pi)^3} \sum_f \left[(c_v^{f})^2 + (c_a^{f})^2 \right] \int\limits_\beta^1 \dd{z} \left[\left(\frac{\beta}{z}\right)^2 + \left(1-\frac{\beta}{z}\right)^2\right] \int\limits_0^{Q^2} \dd{k^2} k^4 \ln \frac{Q^2}{k^2} \\
       & \qquad\times\int \dd{b} b \left\{ \int \dd{r} r J_2(\sqrt{1-z}kr)K_2(\sqrt{z}kr)[2\Tilde{N}(\xpom, \rt;\bt)]\right\}^2.
        \end{aligned}
    \end{equation}
\end{subequations}
\end{widetext}
Here we have replaced the $q\bar q g$-target scattering amplitude $\Tilde{N}_{\ut\rt\bt}$ by an adjoint dipole amplitude $\tilde{N}(\xpom,\rt;\bt)$ which, in the large-$\nc$ limit, becomes $\tilde{N}(\xpom,\rt;\bt) = 2N(\xpom,\rt;\bt) - N^2(\xpom,\rt;\bt)$. 

\section{Diffractive cross section with event-by-event fluctuations}

\label{app:incoherent_diffraction}%

Including event-by-event fluctuations in the scattering amplitude results in non-zero incoherent cross section, see Eq.~\eqref{eq:incoherent_diffraction}. In this Appendix we calculate the diffractive structure functions, including the $q\bar q$ and $q\bar q g$ contributions, in the case of event-by-event fluctuating nuclear structure. The obtained structure functions correspond to the case where the target nucleus is allowed (but not required) to break up, corresponding to the sum of coherent and incoherent diffractive cross sections.

Fluctuations in the scattering amplitude originate from the event-by-event fluctuating nucleon positions $\{ \bt_i \}$, and the event-by-event dipole-nucleus scattering amplitude reads
\begin{equation}
    N_A^{\left\{\bt_{i}\right\}}(\xpom,\rt;\bt) = 1 - \prod_{i=1}^A \left[1-T_p(\bt-\bt_i)\mathcal{N}(\xbj,r)\right]. 
\end{equation}
Note that $T_p$ is dimensionless and satiesfies the normalization condition of \cref{eq:bk_equation}.
As mentioned in the main text, we only consider the {\em after evol.} setup in this case as we do not solve an impact parameter dependent BK evolution equation in this work. The event-by-event dipole-nucleus amplitude is now not rotationally symmetric. 
The $|t|$ integrals can still be performed analytically. 

The obtained diffractive structure functions corresponding to the $q\bar q$ component are
\begin{widetext}
\begin{subequations}
\label{eq:totaldiff_qq_W_Z}
    \begin{align}
        \xpom F^{D(3);W^{\pm}}_{q\bar q,L} &=  \frac{N_c Q^6}{\pi^3 \beta} \sum_{\rm \rm ff'} |V_{\rm \rm ff'}|^2 \int \dd{z} z^3 (1-z)^3 Z_0, \\
        \xpom F^{D(3);W^{\pm}}_{q\bar q,T} &=  \frac{N_cQ^4}{4\pi^3 \beta} \sum_{\rm ff'} |V_{\rm ff'}|^2 \int_{0}^{1} \dd{z}  z(1-z),  \\
         \bQ^2\left[z^2 + (1-z)^2\right] Z_1,
        \xpom F^{D(3);Z}_{q\bar q;L} &= \frac{N_cQ^6}{2\pi^3 \beta} \sum_{f} \left[(c_v^f)^2 + (c_a^f)^2\right] 
        \int_0^1 \dd{z} z^3(1-z)^3 Z_0, \quad \text{and} \\
        \xpom F^{D(3);Z}_{q\bar q;T} &= \frac{N_cQ^4}{8\pi^3 \beta} \sum_{f} \left[(c_v^f)^2 + (c_a^f)^2\right] 
         \int_{0}^{1} \dd{z}  z(1-z) \bQ^2\left[z^2 + (1-z)^2\right] Z_1.
    \end{align}
\end{subequations}

Similarly, for the $q\bar q g$ contribution we obtain

    \begin{subequations}
    \label{eq:totaldiff_qqg_W_Z}
        \begin{equation}
            \begin{aligned}
                \xpom F_{T,q\bar q g}^{D(3);W^\pm} & = \frac{\alpha_s\beta}{(2\pi)^4} \sum_{\sc ff'}|V_{ff'}|^2 \int\limits_\beta^1 \dd{z} \left[\left(\frac{\beta}{z}\right)^2 + \left(1-\frac{\beta}{z}\right)^2\right] \int\limits_0^{Q^2} \dd{k^2}k^4 \ln \frac{Q^2}{k^2} \\
                & \qquad\times\int \dd[2]\bt  \left\{ 2\int \dd{r} r J_2(\sqrt{1-z}kr)K_2(\sqrt{z}kr)\tilde{N}_A^{\left\{\bt_i\right\}} (\xpom,\bt,\rt)\right\}^2,
            \end{aligned}   
        \end{equation}
        and
        \begin{equation}
            \begin{aligned}
                \xpom F_{T,q\bar q g}^{D(3),Z} & =  \frac{\alpha_s\beta}{4(2\pi)^4} \sum_f \left[(c_v^{f})^2 + (c_a^{f})^2 \right] \int\limits_\beta^1 \dd{z} \left[\left(\frac{\beta}{z}\right)^2 + \left(1-\frac{\beta}{z}\right)^2\right] \int\limits_0^{Q^2} \dd{k^2} k^4 \ln \frac{Q^2}{k^2} \\
                & \qquad\times\int \dd[2]{\bt} \left\{ 2\int \dd{r} r J_2(\sqrt{1-z}kr)K_2(\sqrt{z}kr) \tilde{N}_A^{\left\{\bt_i\right\}} (\xpom,\bt,\rt) \right\}^2.
            \end{aligned}
        \end{equation}
    \end{subequations}
Here
\begin{equation}
    Z_{n} \equiv  \int \dd[2]\bt  \left( \int \dd{r} r J_n (\kappa r )K_n(\bQ r) N_A^{\left\{\bt_i\right\}} (\xpom,\bt,\rt) \right)^2. 
\end{equation}
Note that the dependence on the angle of $\bt$  cannot be integrated out due to the lack of rotational symmetry unlike in \cref{eq:qqg_diff_structurefun}.

\end{widetext}
\end{appendix}

\bibliographystyle{JHEP-2modlong}
\bibliography{refs}

\end{document}